%% file: msO2_allsky.tex
\documentclass[aps, twocolumn, superscriptaddress, showpacs, nofootinbib, longbibliography]{revtex4-1}

\usepackage[utf8]{inputenc}
\usepackage[T1]{fontenc}
\usepackage{ae,aecompl} 
\usepackage{graphicx}
\usepackage{amsmath}
\usepackage{color}
\usepackage{amssymb}
\usepackage{latexsym}
\usepackage{wasysym}
\usepackage{psfrag}
\usepackage{ifthen}
\usepackage[citecolor=blue,colorlinks=true]{hyperref}
\usepackage{longtable}
\usepackage{float}
\usepackage{lineno}

\begin{document}

\title{All-sky search for short gravitational-wave bursts in the second Advanced LIGO and Advanced Virgo run}

  \let\mymaketitle\maketitle
  \let\myauthor\author
  \let\myaffiliation\affiliation
  \author{The LIGO Scientific Collaboration}
  \author{The Virgo Collaboration}
 
\begin{abstract}

We present the results of a search for short-duration gravitational-wave transients in the data from the second observing run of Advanced LIGO and Advanced Virgo. We search for gravitational-wave transients with a duration of milliseconds to approximately one second in the 32-4096 Hz frequency band with minimal assumptions about the signal properties, thus targeting a wide variety of sources.  We also perform a matched-filter search for gravitational-wave transients from cosmic string cusps for which the waveform is well-modeled.
The unmodeled search detected gravitational waves from several binary black hole mergers which have been identified by previous analyses. No other significant events have been found by either the unmodeled search or the cosmic string search.
We thus present search sensitivity for a variety of signal waveforms and report upper limits on the source rate-density as function of the characteristic frequency of the signal.  These upper limits are a factor of three lower than the first observing run, with a $50\%$ detection probability for gravitational-wave emissions with energies of $\sim10^{-9}M_\odot c^2$ at 153 Hz. For the search dedicated to cosmic string cusps we consider several loop distribution models, and present updated constraints from the same search done in the first observing run.

\end{abstract}

\maketitle

\section{Introduction}

The Advanced LIGO and Advanced Virgo detectors \cite{TheLIGOScientific:2014jea, TheVirgo:2014hva} have completed their second observing run (O2) which lasted from November 30, 2016 to August 25, 2017. During O2, gravitational-waves (GWs) were detected from seven binary black hole mergers \cite{O2Catalog}, as well as the first binary neutron star merger ever observed \cite{GW170817}.  While binary systems of compact objects such as black holes and/or neutron stars are a main source of short-duration transient GWs observable by LIGO and Virgo, there are other predicted sources of GW transients.  Some examples include core-collapse supernovae \cite{Fryer:2011zz}, pulsar glitches \cite{PhysRevLett.87.241101}, neutron stars collapsing into black holes \cite{Baiotti_2007}, and cosmic string cusps \cite{DV-2000, DV-2001, PhysRevD.71.063510}. There also exists the possibility of new, as-of-yet unpredicted GW sources.

In order to maximize our ability to detect any such GWs, there exist a variety of so-called all-sky searches-- those with no prior assumption on the time of arrival of the GW signal or its location in the sky.  These searches fall broadly into two categories: searches that target GWs from specific sources, and those that look for GWs using minimal assumptions about the source or signal morphology.  Targeted analyses include searches for merging stellar-mass binary black holes and neutron stars \cite{O2Catalog} as well as intermediate mass black holes \cite{PhysRevD.96.022001}, and searches for cosmic string signals \cite{Abbott:2009rr, Aasi:2013vna, Abbott:2017mem}.  The more generic analyses look for both long-duration GW transients \cite{Abbott:2015vir,Abbott:2017muc,Abbott:2019heg} and short-duration events \cite{Abadie:2010mt, s6Burst, O1AllSky}.
In this paper, we report on the results of two all sky searches.  The first is a generic search for short-duration GW transients. The second is a targeted search for cosmic string signals using the matched filtering method with template waveforms predicted from past theoretical studies \cite{DV-2000, DV-2001, PhysRevD.71.063510}.

The rest of this paper is organized as follows: in Section \ref{Sec: o2run} we review the data set used for these analyses. Section \ref{Sec: unmodeled events} is dedicated to the search for unmodeled GW transients and is divided into three parts. First, in \ref{Subsec: searches}, we describe the three search algorithms used to look for generic unmodeled GW transients and the results of those searches. Second, in \ref{Known BBH Signals} we discuss briefly some aspects regarding the detection of the known BBH signals.
In \ref{Subsec: sensitivity}, we discuss the sensitivity of these searches and give rate-density limits of transient GW events, excluding known compact binary sources. Section \ref{Sec: cosmic strings} is dedicated to the modeled cosmic string cusps search. We briefly outline the search algorithm used for the analysis, and present our results and updated parameter constraints. Finally, in Section \ref{Sec: discussion}, we discuss the results and implications from both the unmodeled GW transients search and the modeled cosmic string cusp search.
\section{O2: The Second Advanced-Detector Observing Run}\label{Sec: o2run}

Our data set ranges from November 30, 2016 to August 25, 2017.
Prior to August 2017, only the Hanford and Livingston Advanced LIGO detectors were in observational mode.
On August 1, 2017, Advanced Virgo joined the detector network.
During O2, the combined Hanford-Livingston network sensitivity was slightly more sensitive than it was in the first bserving run (O1), achieving a roughly $30\%$ increase in binary-neutron-star (BNS) range~\cite{ObservingScenario}.
The Advanced Virgo detector was less sensitive than the Advanced LIGO detectors, with a BNS range that was roughly a factor of 2-3 lower~\cite{ObservingScenario}. As a result of this, including the Virgo data set did not improve the sensitivity to the short-duration searches presented in this paper. We thus present the analysis of only the Hanford-Livingston data.

Over the course of O2, the livetime of the data collected by the two LIGO detectors was about 158 days for Hanford, and about 154 days for Livingston. 
The amount of coincident data between the two detectors is approximately 118 days.
Not all of this data is ultimately analyzed though, as the data can sometimes be polluted by instrumental and environmental noise artifacts.  In particular, transient noise events known as ``glitches''  can potentially mimic GW properties thereby lowering the sensitivity of searches for short-duration GW bursts.
 To mitigate the effect of instrumental and environmental noise, a large number of auxiliary channels within the interferometer are monitored in order to characterize the relation between artifacts in these channels and the GW strain channel.
 This auxiliary channel information is used to identify periods of poor data quality, which is then excluded from the analysis ~\cite{DetcharCompanion, BurstCompanion, O1-Detchar, Nuttall:2018xhi}.  The calibration uncertainties in O2 data for Hanford and Livingston respectively are 2.6\% and 3.9\% in amplitude, and 2.4 and 2.2 degrees in phase \cite{Viets_2018, PhysRevD.96.102001}.
Additionally, for the first time in Advanced LIGO data, methods to substract some well identified sources of noise from the data are used, increasing Hanford's sensitivity by ~10\% ~\cite{Driggers:2018gii}.  While these methods remove many known artifacts, not all glitches are removed. Thus, the pipelines in this paper have been designed to confidently distinguish between real GW signals and instrumental glitches.
% nb specific uncertainties here https://dcc.ligo.org/DocDB/0150/G1800319/001/G1800319-v1_O2C02_UncertaintyReview.pdf

The data used is this paper is part of the O1 Data Release and O2 Data Release through the Gravitational Wave Open Science Center~\cite{Vallisneri_2015}, and can be found at~\cite{GWOSC}.

\section{Unmodeled GW Transients} \label{Sec: unmodeled events}

We describe here the unmodeled search for short duration transient signals. Given the uncertainty and the 
wide spectrum of expected signals, the algorithms are designed to use minimal assumptions on the
expected waveform and consider signals with a duration of a few seconds or less in the frequency range of 32 Hz to 4096 Hz.
This covers a wide parameter space of sources, including GWs from mergers of compact objects such as neutron stars or black holes.  While there exist more narrowly focused searches that target GWs from compact binary systems which are naturally more sensitive to this type of signal~\cite{Usman_2016, alex_nitz_2017_344823, PhysRevD.95.042001}, the unmodeled searches presented here are sensitive to a wider variety of potential sources.
In this work, we identify and then remove the known binary black hole (BBH) sources in our analysis results, in order to focus on searching for previously unidentified transients.  

We use the same three unmodeled analyses that were used in O1 search \cite{O1AllSky}.  By using multiple pipelines we have the ability to independently verify search results.  Additionally, the regions of parameter space where these algorithms are the most sensitive is not the same for every pipeline, and so the combination of the different approaches increases our ability to detect a wide range of signals.
Below we describe the three different algorithms used to search for transient GW events.

\subsection{Searches}\label{Subsec: searches}

\subsubsection{Coherent WaveBurst}\label{search:cWB}

Coherent WaveBurst (cWB) is an algorithm based on the maximum-likelihood-ratio statistic applied to  
power excesses in the time-frequency domain \cite{cwb2g}. This analysis is done by using a
wavelet transform at various resolutions, as to adapt the time-frequency characterization to the signal features.
cWB has been used in the previous LIGO-Virgo searches for
transient signals \cite{Abadie:2010mt, s6Burst, O1AllSky}.

The cWB analysis is split into two frequency bands: low and high frequency.  The triggers are further divided into search bins, similar to how it was done for the O1 analysis.

The low-frequency analysis covers the parameter space ranging from 32 - 1024 Hz, and performs a down-sampling of the data.  
The triggers are divided into two different bins.  The first bin, $LF1$, is polluted by non-stationary
power-spectrum lines and a class of low frequency, short duration glitches known as ``blip'' glitches 
for which there is no specific data quality veto \cite{DetcharCompanion}. These are selected using the 
same criteria described in \cite{BurstCompanion}: non-stationary lines localize more than 80\% of their energy in a 
frequency bandwidth of less than 5 Hz; blip glitches are identified according to their waveform properties
so that their quality factor (Q) is less than 3. The second bin, $LF2$ contains the remaining low frequency triggers.
In the O1 analysis \cite{O1AllSky} there was a third class focusing on events with morphology similar
to compact object binaries-- specifically events that chirped up in frequency. This class is not considered in this work, since the results for a cWB dedicated search for chirping signals is reported in \cite{O2Catalog}. The search in \cite{O2Catalog} 
differs from the one presented here in both post-production thresholds and selection of power excesses in 
time-frequency. The latter is performed in \cite{O2Catalog} favoring time-frequency patterns
 with increasing frequency over time. This feature, in addition to dedicated thresholds, reduces the 
background and increases the sensitivity to compact binary coalescence waveforms. 

The high-frequency analysis uses data in the 1024 - 4096 Hz range and is also divided into two bins.  The first bin, $HF1$, contains triggers with central frequency above 2048 Hz, and events with central frequency in the band 1000 - 1150 Hz for
the period of the run before Jan 22nd, 2017.  The second bin, $HF2$, contains the remaining triggers.  The change in the bin definition pre- and post-Jan 22nd is due to an excess of glitches that were occurring around 1100 Hz between October 2016 and January 2017.  These glitches were identified as originating from length fluctuations in the Hanford detector's output mode cleaner optical cavity, and were successfully mitigated for the remainder of O2 \cite{glitch_alog}.

Periods of poor data quality were removed as described in previous searches for short-duration GW events \cite{S5burst, s6Burst, O1AllSky}.  There is some additional loss of livetime in analyzable data because cWB requires at least 1200 seconds of coincident data per analyzable segment.  The final amount of data analyzed by cWB was 113.9 days.

The cWB analysis is performed by dividing the run into reduced periods of consecutive time epochs (called ``chunks'').  Each chunk is composed of about 5 days of livetime, resulting in 21 chunks in total.  The background distribution of triggers for each individual chunk is calculated by time-shifting the data of one detector with respect to the other detector by an amount that breaks any correlation between detectors for a real signal.  Each chunk was time shifted to give about 500 years of background data, which allows the search to reach the  statistical significance of 1/100 years while allowing for a trial factor of 2 for each of the low and high frequency bands.  Performing the analyses in chunks takes into account fluctuating noise levels of the detectors over the duration of the observing run.

The significance of each trigger found in the real coincident data is then calculated by comparing the coherent network signal-to-noise ratio 
$\eta_c$ \cite{O1AllSky} with the background distribution of the chunk to which it belongs.

\begin{figure}[t!]
  \centering
  \scalebox{0.7}{\includegraphics{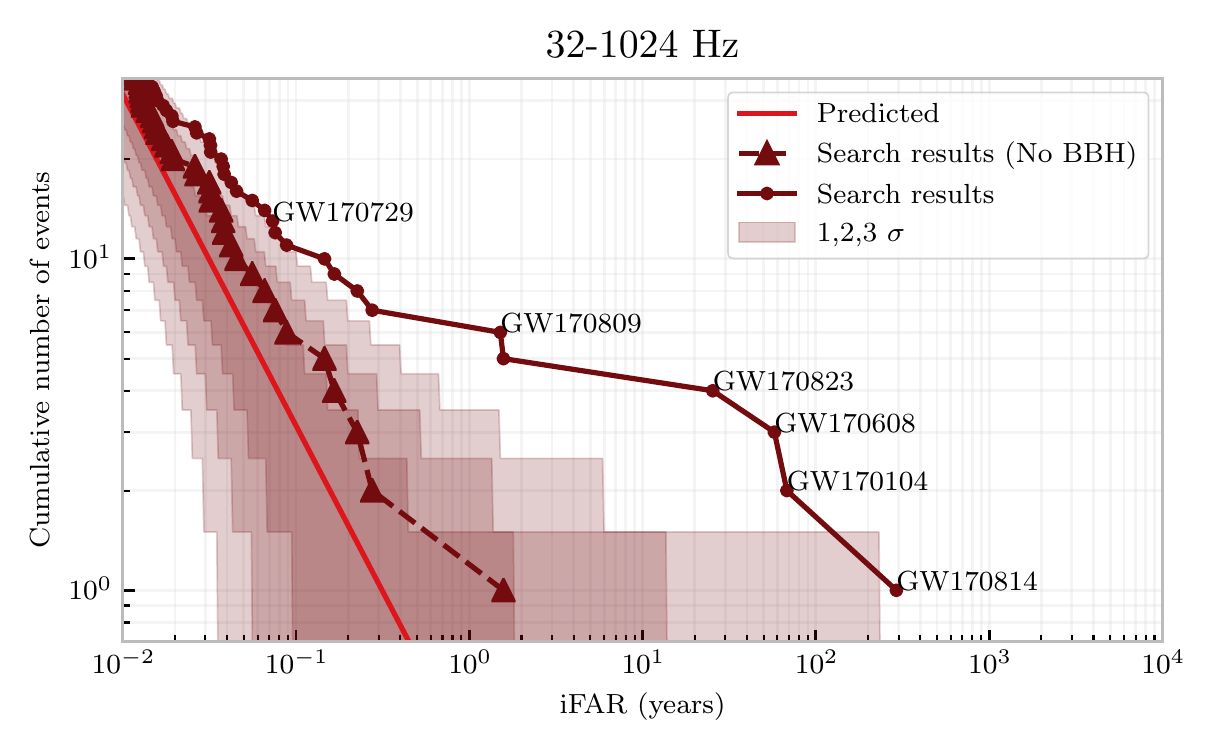}}
  \scalebox{0.7}{\includegraphics{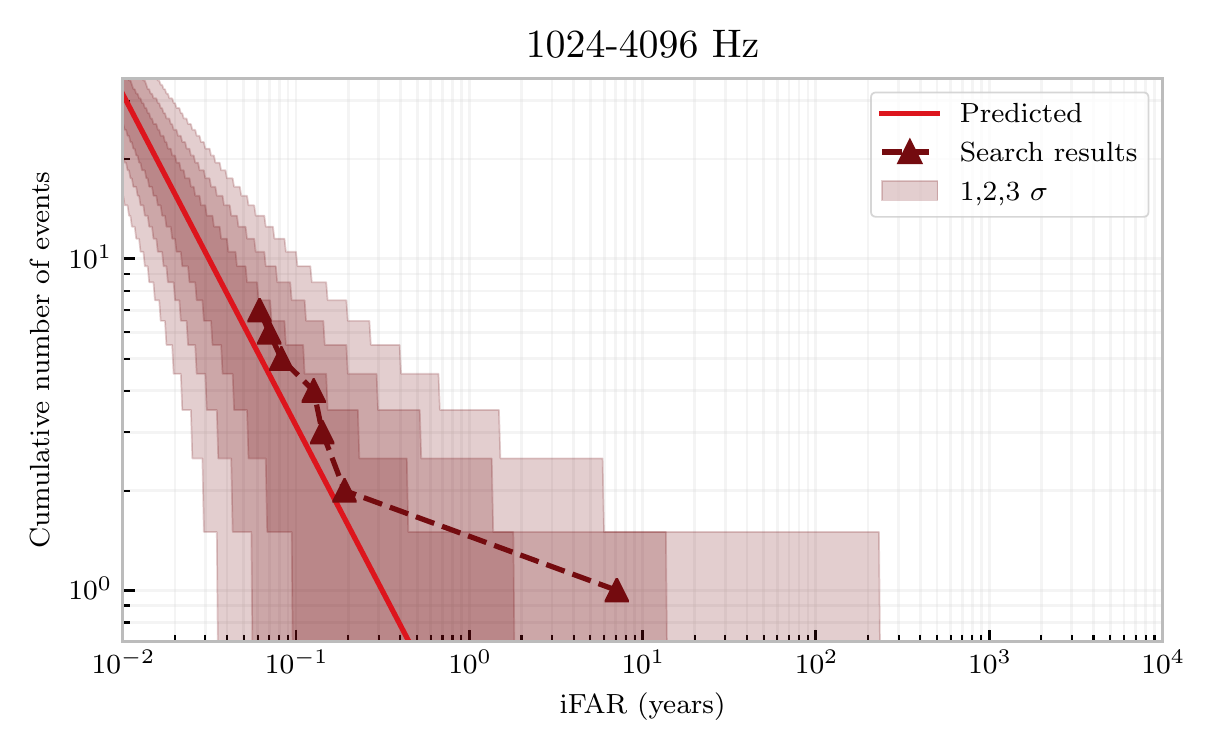}}
  \caption{
  Cumulative number of events versus inverse false alarm rate (iFAR) found by the cWB search using all O2 data (circle points) and the cWB search where times around all compact binary coalescence sources (see table I from \cite{O2Catalog}) have been dropped out (triangular points). The solid line shows the expected background, given the analysis time. 
  The shaded regions show the 1, 2, and 3 $\sigma$ Poisson uncertainty regions.  \emph{Top}: Search results from the cWB low-frequency (32-1024 Hz) band, with results grouped considering all the 
  bins, applying a trials factor equal to 2. 
  \emph{Bottom}: Search results from the cWB high-frequency (1024-4096 Hz) band.  No triggers associated with known BBH signals were found in this search.
  }
  \label{Fig.O2_cWB_FAR}
\end{figure}
The search results for the cWB low and high frequency bands are shown in Fig. \ref{Fig.O2_cWB_FAR}.  In the low frequency search band, cWB found six of the known BBH events with inverse false alarm rates (iFARs) ranging from 290 years for GW170814 
to 0.07 years for GW170729. The loudest trigger in the high-frequency search band has an iFAR of 7 years, and it is related to some disturbances appearing around 1600 Hz.  To search for new events, we remove all previously known GW signals. In this case, this means removing the six BBH signals identified by the search. The remaining events, shown as dashed curves in Fig.~\ref{Fig.O2_cWB_FAR}, are all consistent with expected noise events.

\subsubsection{Omicron-LIB}

Omicron-LIB (oLIB) is a hierarchical search algorithm. oLIB first analyzes the data streams of individual detectors, referred to as an incoherent analysis.  
It then follows up stretches of data that are potentially correlated across the detector network, referred to as a coherent analysis.
The incoherent analysis (``Omicron'')~\cite{OmicronRobinet} flags stretches of coincident excess power.
The coherent follow-up (``LIB'')~\cite{oLIBprep} models GW signals and noise transients with a single sine-Gaussian, and then produces two different Bayes factors.
Each of these Bayes factors is expressed as the natural logarithm of the evidence ratio of two hypotheses: (1) a GW signal versus Gaussian noise (BSN) and (2) a coherent GW signal versus incoherent noise transients (BCI).
The joint likelihood ratio of these two Bayes factors, $\Lambda$, is used as a ranking statistic to assign a significance to each event.

For this analysis, oLIB analyzes two frequency bands:  a low-frequency search band covering 32 - 1024 Hz, and a high-frequency search band covering 1024 - 2048 Hz. 
Similarly to how the analysis was done in O1, low-frequency oLIB event candidates are divided by the quality factor of the signal into high-Q and low-Q search bins (see~\cite{O1AllSky}).
These bins are defined by slightly different cuts than in O1, with the exact choices being made after the background data is analyzed and prior to the analysis of real coincident data.
The low-Q bin contains only events whose median quality factor $\tilde{Q}$ lies within the range of 0.2 - 1.2 and whose median frequency $f_0$ lies within the range of 32 - 1024 Hz. 
The high-Q bin contains only events whose $\tilde{Q}$ lies within the range 2 - 108 and whose $f_0$ lies within the range of 120 - 1024 Hz.  The Q range of 1.2 - 2 is excluded from the analysis \emph{a priori} as that region of parameter space is known to be populated by the blip glitches.
The high-frequency search band contains only events whose $\tilde{Q}$ lies within the range of 2 - 108 and whose $f_0$ lies within the range of 1124 - 2048 Hz.  The lower frequency cut off here is set to 1124 Hz in order to reject a high number of glitches in the 1024-1124 Hz frequency range which were described in~\ref{search:cWB}.
In all bins, event candidates are also required to have positive Bayes factors, 
meaning the GW signal model is favored over the noise models.
A trials factor of 2 is applied to the low-frequency search to account for the independent bins.

Two improvements are made to the O2 oLIB search, as compared to the O1 search that increase the sensitivity.
The first is that $\log \text{BSN}$ is used as a search statistic instead of BSN, which improves the accuracy of oLIB's kernel-density estimates of the signal and noise likelihoods.
Second, event candidates are required to have non-extreme SNR balance across the detector network.
Specifically, we require event candidates to satisfy $\max\lbrace \text{BSN}_\text{H1}/\text{BSN}_\text{L1},\text{BSN}_\text{L1}/\text{BSN}_\text{H1} \rbrace < 9$, where $\text{BSN}_i$ is the BSN Bayes factor estimated using only the data of detector $i$.
This cut helps mitigate the contamination of coincident non-Gaussian noise transients, which tend to have much larger SNR imbalance than GW signals.

\begin{figure}[t!]
  \centering
  \scalebox{0.7}{\includegraphics{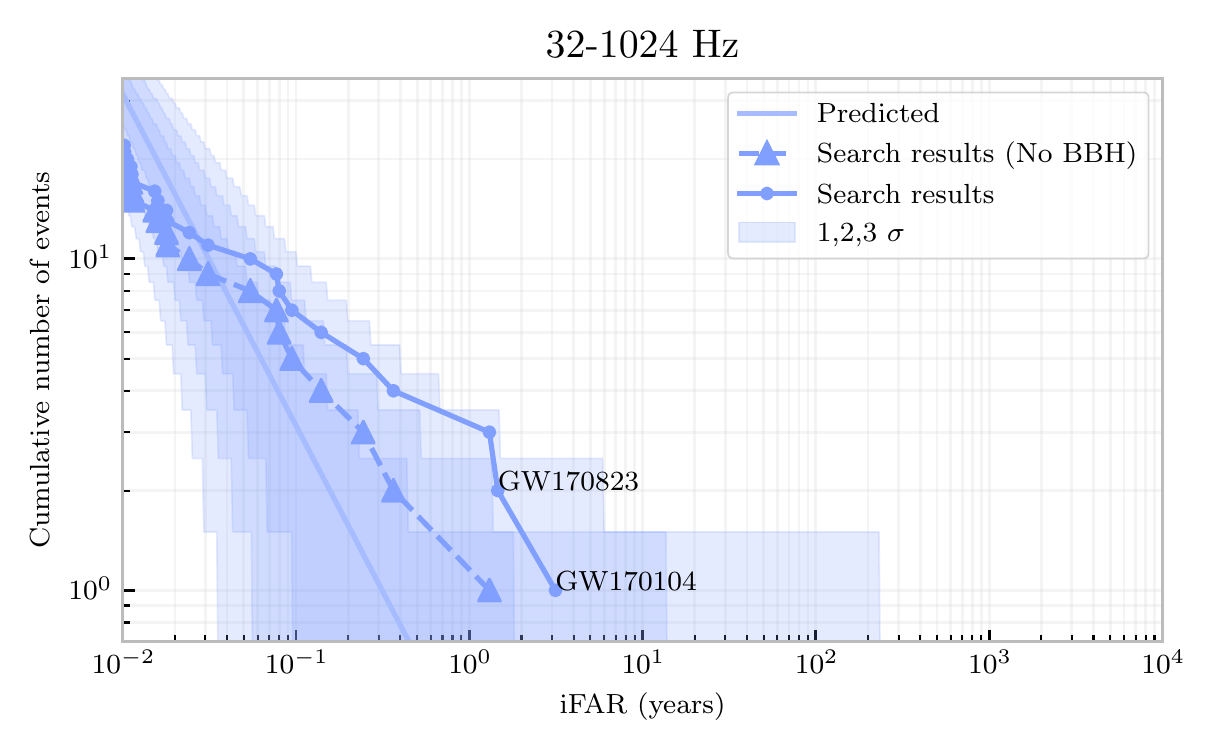}}
  \scalebox{0.7}{\includegraphics{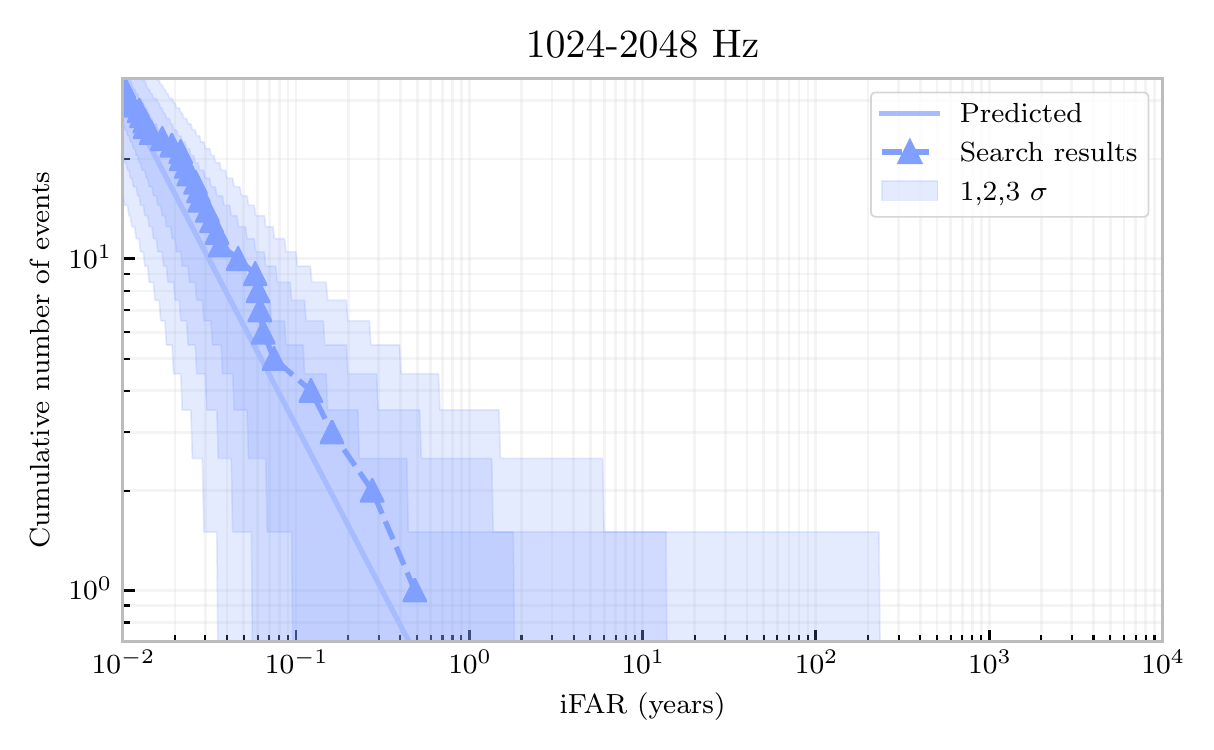}}
  \caption{
  Cumulative number of events versus inverse false alarm rate (iFAR) found by the oLIB search using all O2 data (circle points) and the oLIB search where times around all compact binary coalescence sources (see table I from \cite{O2Catalog}) have been dropped out (triangular points). The solid line shows the expected background, given the analysis time. 
  The shaded regions show the 1, 2, and 3 $\sigma$ Poisson uncertainty regions.  \emph{Top}: The results of the low-frequency (32-1024 Hz) band.  The low-frequency band contains two search bins:  a high-Q bin and a low-Q bin, but as there were no foreground triggers in the low-Q bin, only the high-Q bin is represented here.
  \emph{Bottom}: The search results for the high-frequency (1024-2048) band, which contains only a single search bin.   }
  \label{Fig.O2_oLIB_FAR}
\end{figure}

After removing the periods of poor data quality, oLIB analyzed 114.7 days of coincident detector livetime.  This is slightly more than what was analyzed by cWB because oLIB does not have the same requirement of 1200 seconds of continuous data.  Using the time-slide method, oLIB collected 496 years worth of data to determine the background distribution of glitches.  The significance of triggers found in the zero lag data is calculated by comparing oLIB's ranking statistic to that of the background distribution.
Similar to the O1 analysis, we select single-detector events with SNR > 5.0. The search results are shown in Fig. \ref{Fig.O2_oLIB_FAR}.  
No coincident events satisfy the cuts of the low-Q bin, and the event rate of the high-frequency search matches the expected rate of accidental noise coincidences. 
Two events in the high-Q bin are previously identified BBH events (GW170823 and GW170104).
Again, to search for previously unidentified GW events, the previously known events are removed. 
The results after removing these events are shown as the dashed lines in Fig.~\ref{Fig.O2_oLIB_FAR}.
We notice a small deviation of the high-Q bin's event rate from the expected noise rate for the loudest event candidates, even after all known BBH events are excised from the analysis.
After applying the trials factor of 2, the iFAR of our loudest event candidate is about 1.4 years, which corresponds to a p-value of 0.22.
Using a 5-threshold Event Stacking Test~\cite{EventStacking}, the deviation peaks in significance at the 5$^\mathrm{th}$-loudest event, and the overall p-value of the test is 0.17.
Both of these p-values correspond to one-sided outliers that are less than $1\sigma$ in units of Gaussian standard deviations, and neither signifies a confident detection of GWs.
Thus, we conclude that the oLIB search did not find any new GW events.

\subsubsection{BayesWave Follow-up}

The BayesWave algorithm \cite{Cornish:2014kda, bw_repo} models non-Gaussian features in GW detector data as the sum of sine-Gaussian wavelets using a reversible jump Markov chain Monte Carlo (RJMCMC), where the number of wavelets used is not fixed \emph{a priori} but determined via the RJMCMC.  BayesWave reconstructs the data in two different models: the signal model which treats the data in each interferometer as Gaussian noise plus a common astrophysical signal, and the glitch model which treats the data as Gaussian noise plus independent transient noise artifacts in each detector.  BayesWave then calculates the natural log of the Bayesian evidence of each model.  

The detection statistic used is the log signal-to-glitch Bayes factor ($\ln\mathcal{B}_{sg}$), which is the difference between the logarithm of the two evidences.  A negative $\ln\mathcal{B}_{sg}$ indicates more evidence for a glitch, and a positive $\ln\mathcal{B}_{sg}$ indicates more evidence for a signal.  Beyond minor improvements to the algorithm, the most notable change to BayesWave's mode of operation between O1 and O2 is the prior on the number of wavelets ($N_w$) used in the reconstruction.  While O1 used a flat distribution of $N_w\in[0,20]$ \cite{Cornish:2014kda}, for O2 a prior based on the posterior distribution of $N_w$ during O1 was implemented into the code.  To construct the prior we used the \emph{maximum a posteriori} number of wavelets from a sample of significant background events from O1 to infer the distribution of wavelet dimension. This histogram was then fit to a ratio of polynomials to predict the density at model sizes larger than the O1 cutoff of $N_w=20$. This prior peaks at $N_w=3$, and falls off for higher numbers of wavelets.

In both O1 and O2 BayesWave was used as a follow-up to the cWB pipeline, as adding this follow-up has been shown to enhance confidence in GW detections \cite{Kanner:2015xua}.   For O2, BayesWave followed up cWB events in the low frequency search, treating the $LF1$ and $LF2$ search bins as a single bin, and using a threshold of $\eta_c=9$.  BayesWave used the same approach used by cWB to divide the 113.9 days of analyzable data into chunks of approximately 5 days, and used the same background data set from time slides.

\begin{figure}[t!]
  \centering
  \scalebox{0.7}{\includegraphics{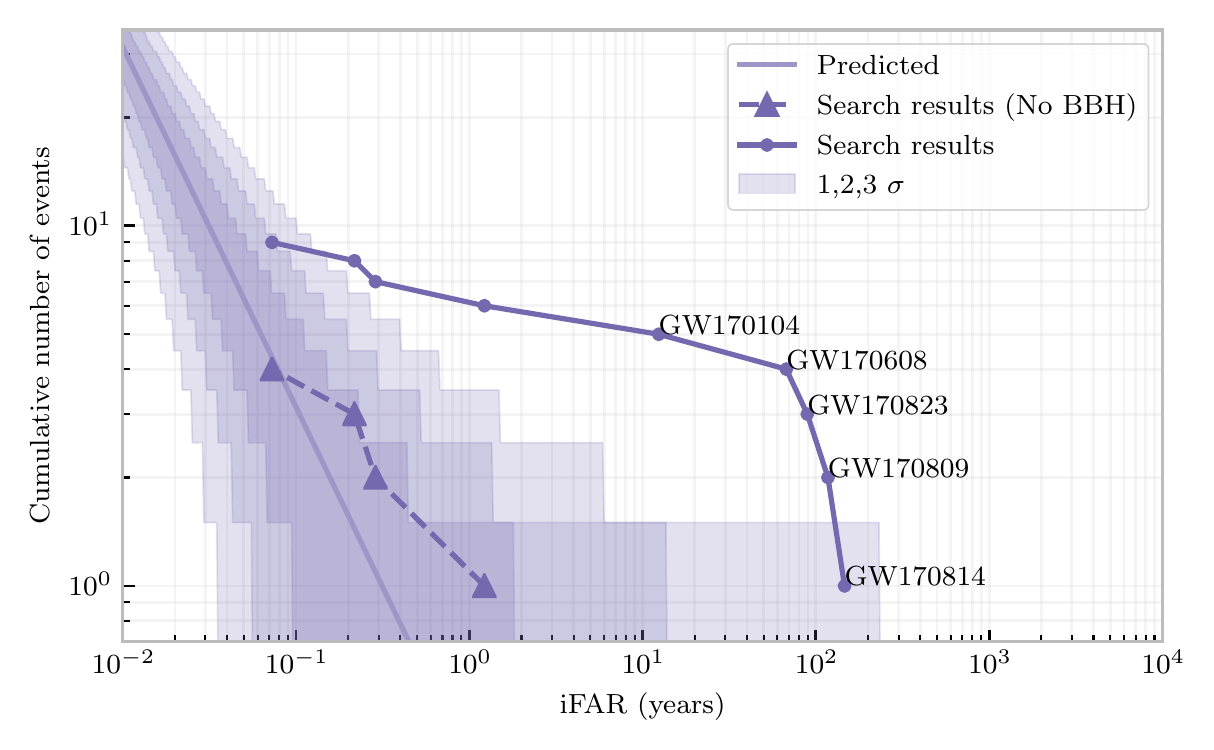}}
  \caption{
  Cumulative number of events versus inverse false alarm rate (iFAR) found by the BW followup to the cWB low-frequency search using all O2 data (circle points) and the BW followup where times around all compact binary coalescence sources (see table I from [3]) have been dropped out (triangular points). The solid line shows the expected background, given the analysis time.
  The shaded regions show the 1, 2, and 3 $\sigma$ Poisson uncertainty regions.
  }
  \label{Fig.O2_BW_FAR}
\end{figure}

There were nine cWB triggers which were above the $\eta_c$ threshold, five of which are known BBH signals\footnote{The only known BBH signal detected by the cWB all-sky algorithm that did not pass the $\eta_c$ threshold was GW170729}.  The results of the BayesWave analysis is shown in Fig. \ref{Fig.O2_BW_FAR}.  The five BBH events were the most significant triggers in the BayesWave results, and after removing them as we did for the cWB and oLIB analysis, all events are consistent with accidental noise fluctuations.

\subsection{Known BBH Signals}\label{Known BBH Signals}
The LIGO and Virgo Collaboration recently released the First GW Transient Catalog (GWTC-1)~\cite{O2Catalog}, which reports all GWs detected by searches targeting compact binary signals in O1 and O2.  GWTC-1 includes ten signals from binary black hole (BBH) mergers, seven of which occurred during O2. These BBHs tend to be short-duration signals that are within the parameter space covered by the unmodeled searches presented here.  So while this search does not target BBH signals, we still found a number of previously identified BBH signals.

Of the seven BBH events in O2, six were identified by at least one of the generic transient search algorithms.  cWB identified six of the BBH events found in O2.  Of those six, five were above the threshold used by the BW followup.  After applying the selection cuts described above, oLIB identifies two of the BBH events-- GW170104 and GW170823. Two other BBH signals, GW170814 and GW170608, are both excluded from the oLIB analysis as a result of narrowly missing some of the data-quality cuts chosen \emph{a priori} for the analysis, but both become clear detections if they are manually added back into the analysis.  One BBH event, GW170818, was not detected by any of the unmodeled pipelines.  The matched filter search in \cite{O2Catalog} that identified GW170818 found it only had an SNR of 4.1 in the Hanford detector.  As the unmodeled analyses are less sensitive to quieter signals like this one, it was missed by this search.
% count the events: 170104, 170608, 170729, 170809, 170814, 170818, 170823

Two cases worth mentioning are GW170729 and GW170809. GW170729 has a lower iFAR than the one given in GWTC-1 \cite{O2Catalog} (50 years). This is expected since, as already explained in Section \ref{search:cWB}, the cWB results reported in GWTC-1 are from a version of cWB with settings for a dedicated for compact binary coalescence search. GW170809 instead was
 not found by cWB in GWTC-1 because that particular time-frequency selection included 
noise excesses. This decreases the coherence of this event between the detectors, which means 
it did not pass one of the post-production thresholds and thus was not assigned any significance.

There was also one binary neutron star merger (GW170817) detected in O2 \cite{GW170817}.  This was a longer signal than the BBH events, appearing in the LIGO data for almost 30 seconds.  The unmodeled pipelines presented here search for signals with a duration of about one second or less, and so did not detect GW170817.

We defer discussion of the astrophysical properties and implications of these events to GWTC-1.  
For the remainder of this paper, we excise known BBH events from our results and place upper-limits on event rates from sources that have not been previously identified by targeted search pipelines.

\subsection{Sensitivity} \label{Subsec: sensitivity}

We measure the detection efficiency of the searches for unmodeled transient events by adding simulated GW signals into real detector data, and using the unmodeled analyses described in \ref{Subsec: searches} to search for these injected signals.  In this work, we use as a detection threshold an iFAR of 100 years.

We do not have accurate waveforms for many of the potential sources in the parameter space of the unmodeled analyses described here.
However, a variety of waveform morphologies can be used to approximate physical situations that are likely to be generated by astrophysical systems. We use these waveforms, distributed through a wide range of amplitudes, durations, and characteristic frequencies to test our unmodeled searches.

\subsubsection{Injection Data Set}

The set of injected signals used in this analysis includes sine-Gaussian (SG), Gaussian (GA), and white-noise burst (WNB) waveforms.  These waveforms, which are not derived from any particular astrophysical model, are the standard in the testing and development of searches for unmodeled GW signals \cite{O1AllSky,s6Burst}.  Each of these injected waveforms can be described by a few characteristic parameters: SG waveforms are parameterized by their central frequency ($f_0$) and quality factor ($Q$); GA waveforms are parameterized by the duration ($\tau$); and finally WNB waveforms are parameterized by their bandwidth ($\Delta f$), lower frequency bound ($f_\mathrm{low}$), and duration in time ($\tau$).  Details about the specifics of these waveforms can be found in \cite{s6Burst}.  To fully test the pipelines sensitivity to range of signals, these waveforms are injected with a range of amplitudes, which we measure as the root-mean-square strain ($h_\mathrm{rss}$) of the waveform at earth.

The injected signal set for this work was produced using \texttt{Minke} \cite{daniel_williams_2018_1699336}, an open-source Python package developed during the O1 detector run. It produces data that contains simulated transient GW signals using the signal generation provided by \texttt{LALSimulation} routines as a part of the LIGO Algorithm Library~\cite{lalsuite}.

For the signal set used in this analysis, signals were produced at a rate of once every 50 seconds.
These were spaced evenly throughout the total time of the run, although the centre time of each signal is shifted by a time drawn from a uniform distribution, between -5 s and +5 s from each division of the timespan. The $h_\mathrm{rss}$ of each signal was drawn from the distribution $r+50/r$, which is uniform in the square of signal distance $r^2$, constructed such that the minimum  $h_\mathrm{rss}$ produced was $5 \times 10^{-23}$, and the maximum $1 \times 10^{-20}$.

Signals are produced for each of the detectors, with the sky location chosen by drawing from a uniform distribution across the sky, and a uniform distribution over waveform polarization; the waveform's sky location is used to calculate the injection time for each signal for each detector. The remaining parameters of each waveform are held fixed for each injection set. 

\begin{table}[!htb]
\centering
\scriptsize
\begin{tabular}{|c||c|c|c|}
\hline
\textbf{Morphology} & \textbf{cWB} & \textbf{oLIB} & \textbf{BW} \\
\hline
\textbf{Gaussian pulses}\\
\hline
$\tau = 0.1$ ms & 8.4 & 6.2 & N/A \\
\hline
$\tau = 2.5$ ms & 11 & 5.3 & N/A \\
\hline
\textbf{sine-Gaussian wavelets}\\
\hline
$f_0 = 70$ Hz, $Q = 3$ & 4.9 & - & N/A \\
\hline
$f_0 = 70$ Hz, $Q = 100$ & 6.4 & - & N/A \\
\hline
$f_0 = 153$ Hz, $Q = 8.9$ & 1.4 & 1.3 & 16 \\
\hline
$f_0 = 235$ Hz, $Q = 100$ & 3.3 & 1.1 & 1.4 \\
\hline
$f_0 = 554$ Hz, $Q = 8.9$ & 1.8 & 1.5 & N/A \\
\hline
$f_0 = 849$ Hz, $Q = 3$ & 5.5 & 2.0 & 17 \\
\hline
$f_0 = 1304$ Hz, $Q = 9$ & 3.3 & 2.8 & - \\
\hline
$f_0 = 1615$ Hz, $Q = 100$ & 3.6 & 3.3 & - \\
\hline
$f_0 = 2000$ Hz, $Q = 3$ & 5.4 & 5.3 & - \\
\hline
$f_0 = 2477$ Hz, $Q = 8.9$ & 7.5 & - & - \\
\hline
$f_0 = 3067$ Hz, $Q = 3$ & 9.7 & - & - \\
\hline
\textbf{White-Noise Bursts}\\
\hline
$f_{low} = 100$ Hz, $\Delta f = 100$ Hz, $\tau = 0.1$ s & 1.4 & 3.0 & 3.0 \\
\hline
$f_{low} = 250$ Hz, $\Delta f = 100$ Hz, $\tau = 0.1$ s & 1.4 & 3.8 & 3.8 \\
\hline
$f_0 = 750$ Hz, $\Delta f = 100$ Hz, $\tau = 0.1$ s & 1.8 & 3.7 & 4.2 \\
\hline
\end{tabular}
\caption{The $h_\text{rss}$ values, in units of $10^{-22} \text{Hz}^{-1/2}$, at which 50\% detection efficiency is achieved at a FAR of 1 in 100 yr for each of the algorithms, as a function of the injected signal morphologies.  ``N/A'' denotes that 50\% detection efficiency was not achieved. ``-'' denotes the waveform was not analyzed by oLIB and BW because its characteristic frequency did not meet the search cuts.}\label{Tab.O2hrss50}
\end{table}

\subsubsection{Results}

Table~\ref{Tab.O2hrss50} shows the specific parameters of all the waveforms analyzed here, and the $h_\text{rss}$ value at which 50\% of the injections are detected by each pipeline for each signal morphology.
The O2 search is more sensitive than in O1.
This increase in efficiency can be attributed to both the increase in detector sensitivity and the improvements made to the algorithms to better deal with instrumental noise.

The introduction of analysis in chunks, for instance, allows for adapting the threshold to the level
of nearby background noise. Moreover, cWB is now using two search bins instead of three. 
Consequently the threshold value
applied to $\eta_c$ decreases at the same FAR. 
The combination of the two effects leads to significant improvements in the efficiency for waveforms belonging 
to the $LF1$ bin with respect to O1 results.

oLIB cuts and tunings are especially beneficial for the GA and WNB waveforms, as oLIB now achieves $50\%$ detection efficiency for all of these waveform morphologies, which it did not achieve in O1.
Nevertheless, these additional cuts do hurt the detection efficiency in some regions of parameter space, such as the band below 120 Hz in the high-Q bin.
For example, the detection efficiency of the SG waveform at 70 Hz is exactly 0 (although oLIB's detection efficiency for this morphology was also negligible in O1 due to its long $\sim 1.5$ s duration).

The BayesWave followup is the least sensitive to SG signals, as shown in \cite{Kanner:2015xua}.  BayesWave's detection statistic, $\ln\mathcal{B}_{sg}$ scales linearly with the number of sine-Gaussian basis functions used in the signal reconstruction, meaning for simple signals that can be accurately represented with a single sine-Gaussian it is harder to distinguish between the signal and glitch models~\cite{Littenberg:2015kpb}.  For signals with more complicated structure in time-frequency space (such as BBH signals which increase in frequency over time), BayesWave is more efficient at distinguishing between the signal and glitch models.  Since the SG and GA waveforms used here can be accurately modeled as a single sine-Gaussian wavelets, BayesWave is less sensitive to these signals.
One improvement made between O1 and O2 is the addition of a jump proposal in the MCMC that helped with the mixing of higher $Q$ signals. This resulted in an increased sensitivity to higher $Q$ signals.

\begin{figure}[t!]
    \centering
    \scalebox{0.9}{\includegraphics[width=0.5\textwidth]{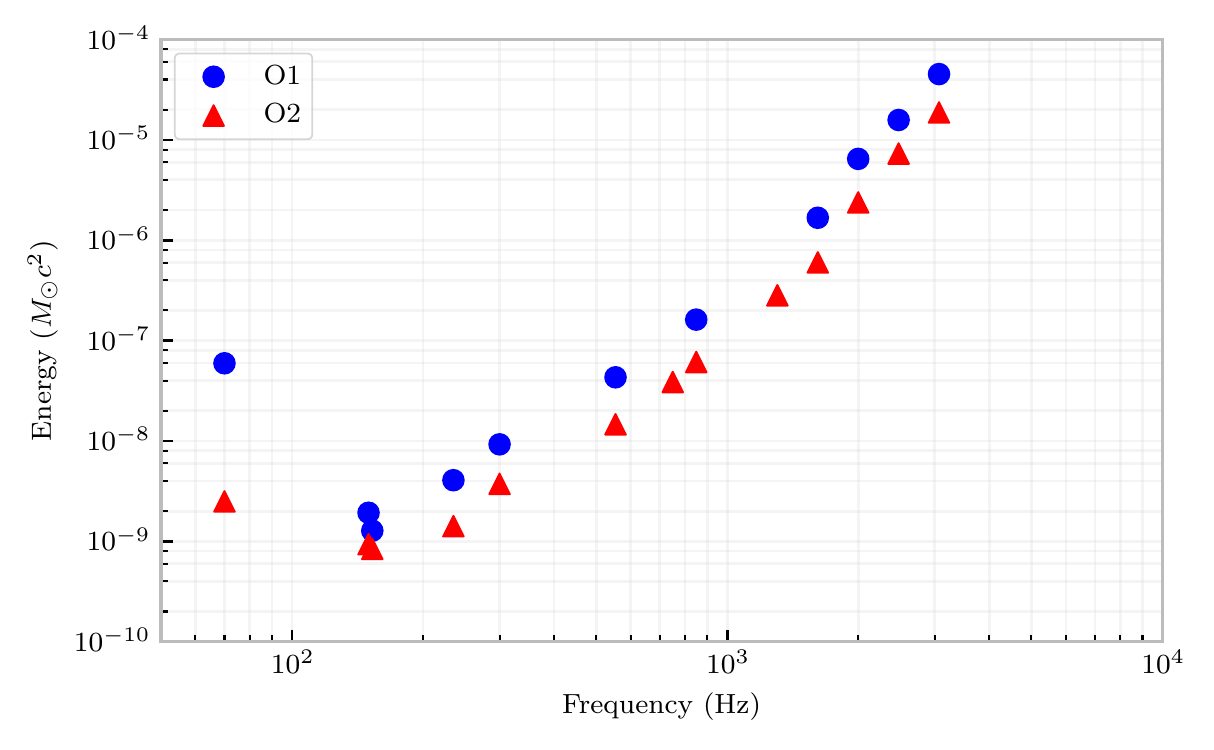}}
    \caption{
    The GW emitted energy in units of solar masses that correspond to a 50\% detection efficiency at an iFAR of 100 years, for a source emitting at 10 kpc.  The waveforms represented here include all of the sine-Gaussian and white noise burst injections as give in Table \ref{Tab.O2hrss50}.   We present the best sensitivity achieved by any of the unmodeled search pipelines, for both the O1 \cite{O1AllSky} and O2 searches.
    }
    \label{Fig.O2Evf}
\end{figure}

From the detection efficiencies given in Table \ref{Tab.O2hrss50}, we can make a statement on the minimum amount of energy emitted by a GW to be detected.  To do this, we assume a standard candle source at a distance of $r_0=10 \mathrm{kpc}$ radiating GWs at a central frequency of $f_0$.  The amount of energy radiated is then \cite{s6Burst}
\begin{equation}
  \label{Eq:Egw}
  E_{GW} = \frac{\pi^2 c^3}{G}r_0^2 f_0^2 h_0^2.
\end{equation}
We use the $h_\mathrm{rss}$ values of 50\% detection efficiency given in Table \ref{Tab.O2hrss50} to find the minimum amount of energy that needs to be radiated by the GW source in order to be detected by at least one of the unmodeled searches.  These results are shown in Fig.~\ref{Fig.O2Evf}, along with the results from the O1 unmodeled all sky search \cite{O1AllSky} for comparison.

\begin{figure}[t!]
    \centering
    \scalebox{0.9}{\includegraphics[width=0.5\textwidth]{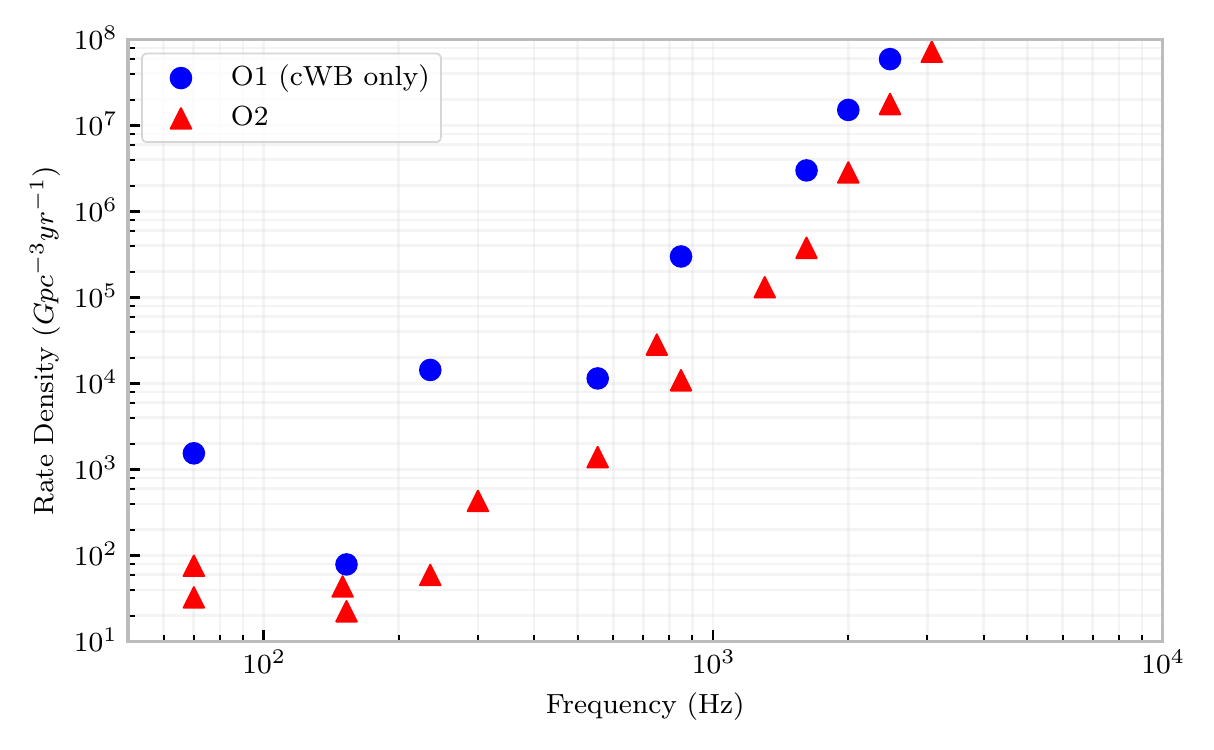}}
    \caption{
    Upper limits on the  90\% confidence intervals for the GW rate-density, as measured in O2 using the SG and WNB waveforms listed in Table~\ref{Tab.O2hrss50}.  Here we show the strictest upper limit achieved by any of the three unmodeled search pipelines.  These results can be scaled to any emission energy $E_{GW}$ using $\textit{rate-density} \propto E_{GW}^{-3/2}$. We also show the results from the O1 All Sky search \cite{O1AllSky}, which presented results from the cWB pipeline for sine-Gaussian waveforms. Note that O1 cWB search 
was using three bins, mostly affecting the efficiency for waveforms belonging to LF1 (i.e. 70 and 235 Hz blue
dots).
    }
    \label{Fig.O2UpperLimits}
\end{figure}

Given that the searches did not find any additional detection results for GW sources beyond the known BBH signals, we can update the upper limit of the rate per unit volume of non-BBH standard-candle sources~\cite{s6Burst,O1AllSky}, shown in Fig.~\ref{Fig.O2Evf}. 
For these upper limits, we use the SG and WNB injection sets listed in Table~\ref{Tab.O2hrss50} as representative morphologies of non-BBH GW bursts.
The markers represent the upper limit at 90\% confidence for rate-density~\cite{s6Burst}, calculated assuming that no noise events meet the detection threshold in our analysis data.  The results shown in Fig. \ref{Fig.O2UpperLimits} assume that $1\ M_\odot c^2$ of GW energy has been emitted from the source, but the upper limits can be scaled to any emission energy $E_{GW}$ by using Eq.~\ref{Eq:Egw} to find that the rate-density scales $\propto E_{GW}^{-\frac{3}{2}}$.

Compared to the rate-density upper limits placed in O1 \cite{O1AllSky} using only the cWB analysis on SG injections, the upper limits reported here for the O2 run are at least a factor of 3 stricter than the O1 upper limits, with much greater improvements at certain frequencies. 
We would expect these upper limits to be at least a factor of 2.4 stricter that the O1 upper limits based on the pipelines analyzing a factor of 2.4 more livetime than in the O1 search.  The greatest improvements in the upper limits between O1 and O2 is due to the fact that here we present the strictest upper limit from any of the three algorithms described here, as opposed to the O1 results which only reported the cWB limits.  Because cWB is not necessarily the pipeline with the greatest sensitivity for every frequency, we get substantial improvement from considering results from all pipelines.

The rest of the improvement can be attributed to the more sensitive Hanford-Livingston detector network, and improvements made to the analysis algorithms.
These upper limits are almost 2 orders of magnitude stricter than those set in all of the initial-detector observing runs (i.e. S5 and S6) at a lower iFAR detection threshold of 8 years.

\section{Cosmic String Cusps}\label{Sec: cosmic strings}

Cosmic strings~\cite{vilenkin-shellard} are one-dimensional topological defects thought to be the relics of phase transitions in the early universe. When a cosmic string interacts with another string in two points or with itself, it intercommutes and forms a loop. Cosmic string loops oscillate and form cusps, which are points along the loop with large Lorentz boosts. Cusps are expected to produce powerful bursts of GWs, having distinct signatures. In particular, the waveform is well predicted by theory~\cite{DV-2000,DV-2001,PhysRevD.71.063510}, which offers the possibility to specifically search for these signals in GW data. Here, we report on a template-based analysis designed to search for GW signals from cosmic string cusps in LIGO and Virgo data. In this work we will focus on Nambu-Goto strings~\cite{Goto:1971ce} whose thickness is approximated to be zero, and assume the intercommutation probability equals unity.

\begin{figure}[t!]
  \centering
  \scalebox{0.8}{\includegraphics[width=0.5\textwidth]{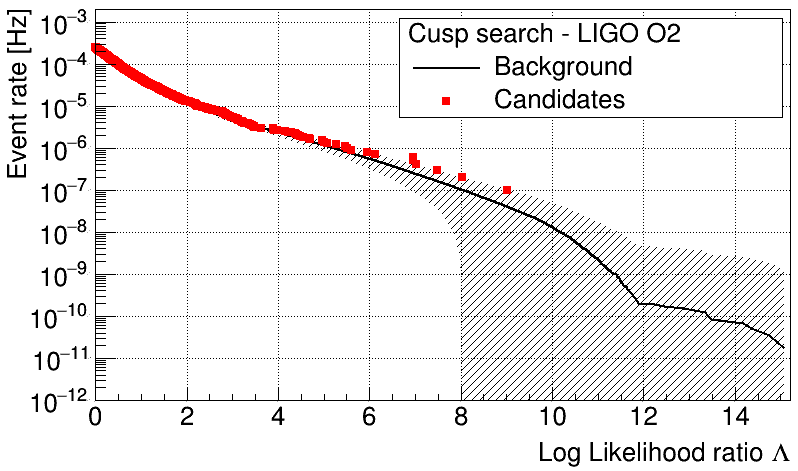}}
  \scalebox{0.8}{\includegraphics[width=0.5\textwidth]{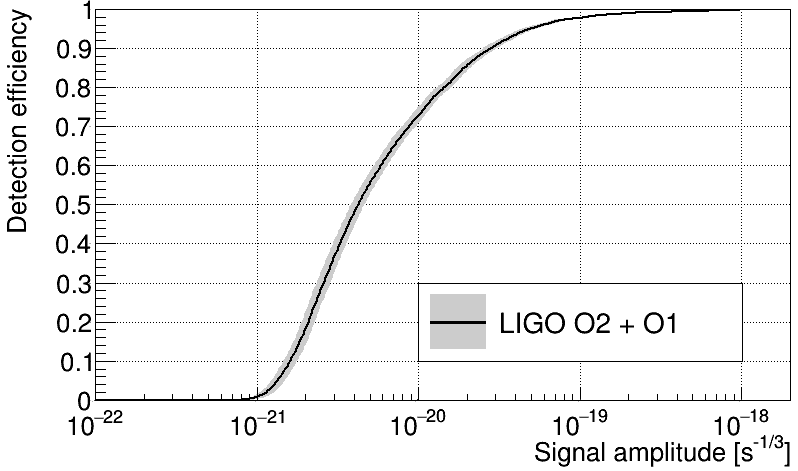}}
  \caption{
    Top: Cumulative event rate for the cosmic string search as a function of the ranking statistic $\Lambda$ (red points). The black line shows the expected background distribution with a $\pm1\sigma $ statistical error represented by the hatched area. The highest-ranked event ($\Lambda_{h}\simeq 9.01$) is consistent with the background. Bottom: search detection efficiency as a function of the cusp signal amplitude, when combining O1 and O2 LIGO data sets. This is measured by the fraction of simulated cusp events recovered with $\Lambda > \Lambda _{h}$.
  }\label{Fig.O2_CS}
\end{figure}

\subsection{The Search}
The cosmic string cusp waveform in the frequency domain is given by $h_{\text{cusp}}(f) = A f^{-4/3}$, where $A$ and $f$ are the signal amplitude and frequency respectively~\cite{DV-2000,DV-2001,PhysRevD.71.063510}. The signal spectrum is limited by a high-frequency cutoff determined by the angle between the beamed emission from the cusp and the observer. This parameter is unknown such that a bank of waveform templates, with different high-frequency cutoff values, is used to perform a matched-filter analysis.

We present the results of the search using the O2 data for GW bursts from cosmic string cusps. This search was conducted in the past using initial LIGO-Virgo data~\cite{Aasi:2013vna} and Advanced LIGO O1 data~\cite{Abbott:2017mem}, and no signal was found. Here, we have used the same analysis methods, which we describe briefly below.

GW bursts from cosmic string cusps are searched by projecting the data onto the bank of templates. Triggered events detected at the Hanford and Livingston detectors are then set in time coincidence in order to reject single-detector noise. Then a multivariate likelihood ratio~\cite{Cannon:2008zz} $\Lambda$ is computed for each of the surviving events, and is used to rank the coincident events.

The pipeline's sensitivity to cosmic string signals was estimated by injecting simulated cusp signals of different amplitudes into the data, and checking whether they were recovered by the pipeline. The sensitivity as a function of amplitude can be used to set constraints on cosmic string parameters, if no signals with high significance are found. To maximize the sensitivity to GW signals we performed data quality studies to reject glitches that may mimic the waveform from cosmic string cusps. We have considered all data quality flags and vetoes produced for transient gravitational wave searches ~\cite{s6Burst,DetcharCompanion}. For each chunk, we look at the effectiveness of the data quality flags which is defined to be the ratio of the fraction of glitches removed to the fraction of analyzable lifetime removed by the veto.  We select those for which that ratio is greater than 2.  The search sensitivity was slightly improved by using a selection of data quality flags.

\subsection{Results}
The cumulative event rate as a function of the ranking statistic $\Lambda$ is displayed in the upper plot of Fig.~\ref{Fig.O2_CS}. The highest-ranked event is measured with $\Lambda_{h} \simeq 9.01$. The ranking value of this event does not deviate significantly from the background distribution estimated by performing the analysis over 6000 time-shifted data sets, made by shifting the Livingston data a sufficient amount so that the events surviving the coincidence check will only be noise. Therefore, we cannot claim this event to be the result of a GW signal produced by cosmic strings. Although we cannot rule out the possibility that the highest-ranked events are real signals, data quality studies find that these events are consistent with blip glitches, the time-frequency structure of which matches very well the waveform of a cusp signal. The exact cause of these glitches is unknown.

The lower plot in Fig.~\ref{Fig.O2_CS} shows the detection efficiency of cusp events as a function of the injected signal amplitude $A$. The efficiency is computed as the fraction of simulated signals recovered with $\Lambda > \Lambda _{h}$. Here we show the sensitivity curve combining O1 and O2 data, as the sensitivity of the O2 LIGO cusp search is comparable to the O1 LIGO one~\cite{Abbott:2017mem}.

We also conducted a three-detector search using the data collected by Advanced Virgo in August 2017, corresponding to $\sim 17$ days of data. The search false alarm rate improved, in combination with a significant reduction of coincident blip noise signals. However the sensitivity of the Advanced Virgo detector was not sufficient to improve the detection efficiency of cosmic string signals and so here we only present results using Hanford and Livingston data.

Using the O1 and O2 combined detection efficiency, we place constraints on the string tension $G\mu$ ($c = 1$), where $G$ is the Newton's constant and $\mu$ is the mass per unit length. This is achieved by comparing the experimental sensitivity to cosmic string signals with predicted detection rates. The expected rate can be derived from cosmic string loop distribution models relying on numerical simulations of cosmic string networks. We examine two analytic models of cosmic string loop distributions already used in the O1 analysis. The loop density modeled in~\cite{Blanco-Pillado:2013qja} was first considered. However, the sensitivity to burst signals produced by such loops is not sufficient to constrain $G\mu$ significantly. We tested another loop density modeled in~\cite{Ringeval:2005kr}, where tiny loops are produced in greater amount than in~\cite{Blanco-Pillado:2013qja}, producing a higher rate of GW bursts. In this case, the upper limit on the string tension is $G \mu \leq 4.2\times 10^{-10}$, with a 95\% confidence level. This O1+O2 upper limit has improved by a factor $\sim 2$ with respect to the previous limit obtained with O1 data alone. Under the assumption of these loop distributions, our non-detection is consistent with the non-detection of the stochastic background created by these bursts, from which stronger constraints on $G\mu$ are obtained~\cite{O2stochastic}.

\section{Conclusion}\label{Sec: discussion}

This paper reports the results for two searches for short duration GWs in the second observing run: one for generic unmodeled GW transient signals and the other focused on modeled cosmic string cusps. 

The most generic search for unmodeled GW transients uses  minimal assumptions on the signal waveform, direction or arrival time and is performed using three different methods. 
Apart from the known BBH signals described in detail in \cite{O2Catalog}, no other signals were found by the unmodeled search. We use our null detection to pose rate-density upper limits on short duration transient GW events not associated with BBH systems.
These limits are stricter than the limits derived from the O1 analysis by a factor of at least 3, owing to a combination of better detector sensitivities, increased observation time, and algorithmic developments.

In the search for modeled cosmic string cusps, we select two analytic models for loop distributions already used in the O1 analysis~\cite{Abbott:2017mem}. We improve the constraints on the string tension $G\mu$ for the model that produces a large amount of small loops~\cite{Ringeval:2005kr}. Our results are complemented by the O2 stochastic results~\cite{O2stochastic}, which have obtained tighter constraints on the string tension $G\mu$.  

LIGO and Virgo began their next observing run in April 2019. 
In addition to the detectors already in operation, two new ground-based detectors will join the search for GWs in the future.  KAGRA, in Japan, has just finished installation and is aiming to join the O3 run, and LIGO-India is currently under construction \cite{ObservingScenario}. 
Improved sensitivities and additional detectors  will lead to better sensitivities for short-duration GW searches in the future.

\bigskip\noindent\textit{Acknowledgments} ---
\input{LVCacknowledgements2018}

This article has been assigned the LIGO document number P1800308.

%\bibliography{O2_allsky}
\input{msO2_allsky.bbl}

  \let\author\myauthor
  \let\affiliation\myaffiliation
  \let\maketitle\mymaketitle
  \title{Authors}
  \input{LSC-Virgo-Authors-Feb-2019-prd}
  \newpage
  \maketitle

\end{document}

%% file: LVCacknowledgements2018.tex
The authors gratefully acknowledge the support of the United States
National Science Foundation (NSF) for the construction and operation of the
LIGO Laboratory and Advanced LIGO as well as the Science and Technology Facilities Council (STFC) of the
United Kingdom, the Max-Planck-Society (MPS), and the State of
Niedersachsen/Germany for support of the construction of Advanced LIGO 
and construction and operation of the GEO600 detector. 
Additional support for Advanced LIGO was provided by the Australian Research Council.
The authors gratefully acknowledge the Italian Istituto Nazionale di Fisica Nucleare (INFN),  
the French Centre National de la Recherche Scientifique (CNRS) and
the Foundation for Fundamental Research on Matter supported by the Netherlands Organisation for Scientific Research, 
for the construction and operation of the Virgo detector
and the creation and support  of the EGO consortium. 
The authors also gratefully acknowledge research support from these agencies as well as by 
the Council of Scientific and Industrial Research of India, 
the Department of Science and Technology, India,
the Science \& Engineering Research Board (SERB), India,
the Ministry of Human Resource Development, India,
the Spanish  Agencia Estatal de Investigaci\'on,
the Vicepresid\`encia i Conselleria d'Innovaci\'o, Recerca i Turisme and the Conselleria d'Educaci\'o i Universitat del Govern de les Illes Balears,
the Conselleria d'Educaci\'o, Investigaci\'o, Cultura i Esport de la Generalitat Valenciana,
the National Science Centre of Poland,
the Swiss National Science Foundation (SNSF),
the Russian Foundation for Basic Research, 
the Russian Science Foundation,
the European Commission,
the European Regional Development Funds (ERDF),
the Royal Society, 
the Scottish Funding Council, 
the Scottish Universities Physics Alliance, 
the Hungarian Scientific Research Fund (OTKA),
the Lyon Institute of Origins (LIO),
the Paris \^{I}le-de-France Region, 
the National Research, Development and Innovation Office Hungary (NKFIH), 
the National Research Foundation of Korea,
Industry Canada and the Province of Ontario through the Ministry of Economic Development and Innovation, 
the Natural Science and Engineering Research Council Canada,
the Canadian Institute for Advanced Research,
the Brazilian Ministry of Science, Technology, Innovations, and Communications,
the International Center for Theoretical Physics South American Institute for Fundamental Research (ICTP-SAIFR), 
the Research Grants Council of Hong Kong,
the National Natural Science Foundation of China (NSFC),
the Leverhulme Trust, 
the Research Corporation, 
the Ministry of Science and Technology (MOST), Taiwan
and
the Kavli Foundation.
The authors gratefully acknowledge the support of the NSF, STFC, MPS, INFN, CNRS and the
State of Niedersachsen/Germany for provision of computational resources.

%% file: msO2_allsky.bbl

%% file: LSC-Virgo-Authors-Feb-2019-prd.tex
%% LSC authorlist in PRD format
%\documentclass[aps,prd,superscriptaddress]{revtex4-1}
%\begin{document}

%\title{LSC February 2019 and Virgo February 2019 author list---LIGO-M1900040\\
%4/26/2019. PRD style}

\author{B.~P.~Abbott}
\affiliation{LIGO, California Institute of Technology, Pasadena, CA 91125, USA}
\author{R.~Abbott}
\affiliation{LIGO, California Institute of Technology, Pasadena, CA 91125, USA}
\author{T.~D.~Abbott}
\affiliation{Louisiana State University, Baton Rouge, LA 70803, USA}
\author{S.~Abraham}
\affiliation{Inter-University Centre for Astronomy and Astrophysics, Pune 411007, India}
\author{F.~Acernese}
\affiliation{Dipartimento di Farmacia, Universit\`a di Salerno, I-84084 Fisciano, Salerno, Italy}
\affiliation{INFN, Sezione di Napoli, Complesso Universitario di Monte S.Angelo, I-80126 Napoli, Italy}
\author{K.~Ackley}
\affiliation{OzGrav, School of Physics \& Astronomy, Monash University, Clayton 3800, Victoria, Australia}
\author{C.~Adams}
\affiliation{LIGO Livingston Observatory, Livingston, LA 70754, USA}
\author{R.~X.~Adhikari}
\affiliation{LIGO, California Institute of Technology, Pasadena, CA 91125, USA}
\author{V.~B.~Adya}
\affiliation{OzGrav, Australian National University, Canberra, Australian Capital Territory 0200, Australia}
\author{C.~Affeldt}
\affiliation{Max Planck Institute for Gravitational Physics (Albert Einstein Institute), D-30167 Hannover, Germany}
\affiliation{Leibniz Universit\"at Hannover, D-30167 Hannover, Germany}
% ...

%\collaboration{The LIGO Scientific Collaboration and the Virgo Collaboration}
%\begin{abstract}
%This is the LSC February 2019 and Virgo February 2019 author list---LIGO-M1900040.
%\end{abstract}

%\maketitle

%\end{document}

%% file: msO2_allsky.bbl
\begin{thebibliography}{51}%
\makeatletter
\providecommand \@ifxundefined [1]{%
 \@ifx{#1\undefined}
}%
\providecommand \@ifnum [1]{%
 \ifnum #1\expandafter \@firstoftwo
 \else \expandafter \@secondoftwo
 \fi
}%
\providecommand \@ifx [1]{%
 \ifx #1\expandafter \@firstoftwo
 \else \expandafter \@secondoftwo
 \fi
}%
\providecommand \natexlab [1]{#1}%
\providecommand \enquote  [1]{``#1''}%
\providecommand \bibnamefont  [1]{#1}%
\providecommand \bibfnamefont [1]{#1}%
\providecommand \citenamefont [1]{#1}%
\providecommand \href@noop [0]{\@secondoftwo}%
\providecommand \href [0]{\begingroup \@sanitize@url \@href}%
\providecommand \@href[1]{\@@startlink{#1}\@@href}%
\providecommand \@@href[1]{\endgroup#1\@@endlink}%
\providecommand \@sanitize@url [0]{\catcode `\\12\catcode `\$12\catcode
  `\&12\catcode `\#12\catcode `\^12\catcode `\_12\catcode `\%12\relax}%
\providecommand \@@startlink[1]{}%
\providecommand \@@endlink[0]{}%
\providecommand \url  [0]{\begingroup\@sanitize@url \@url }%
\providecommand \@url [1]{\endgroup\@href {#1}{\urlprefix }}%
\providecommand \urlprefix  [0]{URL }%
\providecommand \Eprint [0]{\href }%
\providecommand \doibase [0]{http://dx.doi.org/}%
\providecommand \selectlanguage [0]{\@gobble}%
\providecommand \bibinfo  [0]{\@secondoftwo}%
\providecommand \bibfield  [0]{\@secondoftwo}%
\providecommand \translation [1]{[#1]}%
\providecommand \BibitemOpen [0]{}%
\providecommand \bibitemStop [0]{}%
\providecommand \bibitemNoStop [0]{.\EOS\space}%
\providecommand \EOS [0]{\spacefactor3000\relax}%
\providecommand \BibitemShut  [1]{\csname bibitem#1\endcsname}%
\let\auto@bib@innerbib\@empty
%</preamble>
\bibitem [{\citenamefont {Aasi}\ \emph {et~al.}(2015)\citenamefont {Aasi} \emph
  {et~al.}}]{TheLIGOScientific:2014jea}%
  \BibitemOpen
  \bibfield  {author} {\bibinfo {author} {\bibfnamefont {J.}~\bibnamefont
  {Aasi}} \emph {et~al.} (\bibinfo {collaboration} {LIGO Scientific
  Collaboration}),\ }\bibfield  {title} {\enquote {\bibinfo {title} {{Advanced
  LIGO}},}\ }\href {\doibase 10.1088/0264-9381/32/7/074001} {\bibfield
  {journal} {\bibinfo  {journal} {Class. Quant. Grav.}\ }\textbf {\bibinfo
  {volume} {32}},\ \bibinfo {pages} {074001} (\bibinfo {year} {2015})},\
  \Eprint {http://arxiv.org/abs/1411.4547} {arXiv:1411.4547 [gr-qc]}
  \BibitemShut {NoStop}%
%%CITATION = ARXIV:1411.4547;%%
\bibitem [{\citenamefont {Acernese}\ \emph {et~al.}(2015)\citenamefont
  {Acernese} \emph {et~al.}}]{TheVirgo:2014hva}%
  \BibitemOpen
  \bibfield  {author} {\bibinfo {author} {\bibfnamefont {F.}~\bibnamefont
  {Acernese}} \emph {et~al.} (\bibinfo {collaboration} {Virgo Collaboration}),\
  }\bibfield  {title} {\enquote {\bibinfo {title} {{Advanced Virgo: a
  second-generation interferometric gravitational wave detector}},}\ }\href
  {\doibase 10.1088/0264-9381/32/2/024001} {\bibfield  {journal} {\bibinfo
  {journal} {Class. Quant. Grav.}\ }\textbf {\bibinfo {volume} {32}},\ \bibinfo
  {pages} {024001} (\bibinfo {year} {2015})},\ \Eprint
  {http://arxiv.org/abs/1408.3978} {arXiv:1408.3978 [gr-qc]} \BibitemShut
  {NoStop}%
%%CITATION = ARXIV:1408.3978;%%
\bibitem [{\citenamefont {Abbott}\ \emph
  {et~al.}(2018{\natexlab{a}})\citenamefont {Abbott} \emph
  {et~al.}}]{O2Catalog}%
  \BibitemOpen
  \bibfield  {author} {\bibinfo {author} {\bibfnamefont {B.~P.}\ \bibnamefont
  {Abbott}} \emph {et~al.} (\bibinfo {collaboration} {LIGO Scientific
  Collaboration and Virgo Collaboration}),\ }\bibfield  {title} {\enquote
  {\bibinfo {title} {{GWTC-1: A Gravitational-Wave Transient Catalog of Compact
  Binary Mergers Observed by LIGO and Virgo during the First and Second
  Observing Runs}},}\ }\href@noop {} {\  (\bibinfo {year}
  {2018}{\natexlab{a}})},\ \Eprint {http://arxiv.org/abs/1811.12907}
  {arXiv:1811.12907 [astro-ph.HE]} \BibitemShut {NoStop}%
%%CITATION = ARXIV:1811.12907;%%
\bibitem [{\citenamefont {Abbott}\ \emph
  {et~al.}(2017{\natexlab{a}})\citenamefont {Abbott} \emph
  {et~al.}}]{GW170817}%
  \BibitemOpen
  \bibfield  {author} {\bibinfo {author} {\bibfnamefont {B.~P.}\ \bibnamefont
  {Abbott}} \emph {et~al.} (\bibinfo {collaboration} {LIGO Scientific
  Collaboration and Virgo Collaboration}),\ }\bibfield  {title} {\enquote
  {\bibinfo {title} {{GW}170817: Observation of gravitational waves from a
  binary neutron star inspiral},}\ }\href {\doibase
  10.1103/PhysRevLett.119.161101} {\bibfield  {journal} {\bibinfo  {journal}
  {Phys. Rev. Lett.}\ }\textbf {\bibinfo {volume} {119}},\ \bibinfo {pages}
  {161101} (\bibinfo {year} {2017}{\natexlab{a}})}\BibitemShut {NoStop}%
\bibitem [{\citenamefont {Fryer}\ and\ \citenamefont
  {New}(2011)}]{Fryer:2011zz}%
  \BibitemOpen
  \bibfield  {author} {\bibinfo {author} {\bibfnamefont {C.~L.}\ \bibnamefont
  {Fryer}}\ and\ \bibinfo {author} {\bibfnamefont {K.~C.~B.}\ \bibnamefont
  {New}},\ }\bibfield  {title} {\enquote {\bibinfo {title} {{Gravitational
  waves from gravitational collapse}},}\ }\href {\doibase 10.12942/lrr-2003-2}
  {\bibfield  {journal} {\bibinfo  {journal} {Living Rev. Rel.}\ }\textbf
  {\bibinfo {volume} {14}},\ \bibinfo {pages} {1} (\bibinfo {year}
  {2011})}\BibitemShut {NoStop}%
%%CITATION = 00222,14,1;%%
\bibitem [{\citenamefont {Andersson}\ and\ \citenamefont
  {Comer}(2001)}]{PhysRevLett.87.241101}%
  \BibitemOpen
  \bibfield  {author} {\bibinfo {author} {\bibfnamefont {N.}~\bibnamefont
  {Andersson}}\ and\ \bibinfo {author} {\bibfnamefont {G.~L.}\ \bibnamefont
  {Comer}},\ }\bibfield  {title} {\enquote {\bibinfo {title} {Probing
  neutron-star superfluidity with gravitational-wave data},}\ }\href {\doibase
  10.1103/PhysRevLett.87.241101} {\bibfield  {journal} {\bibinfo  {journal}
  {Phys. Rev. Lett.}\ }\textbf {\bibinfo {volume} {87}},\ \bibinfo {pages}
  {241101} (\bibinfo {year} {2001})}\BibitemShut {NoStop}%
\bibitem [{\citenamefont {Baiotti}\ \emph {et~al.}(2007)\citenamefont
  {Baiotti}, \citenamefont {Hawke},\ and\ \citenamefont
  {Rezzolla}}]{Baiotti_2007}%
  \BibitemOpen
  \bibfield  {author} {\bibinfo {author} {\bibfnamefont {L.}~\bibnamefont
  {Baiotti}}, \bibinfo {author} {\bibfnamefont {I.}~\bibnamefont {Hawke}}, \
  and\ \bibinfo {author} {\bibfnamefont {L.}~\bibnamefont {Rezzolla}},\
  }\bibfield  {title} {\enquote {\bibinfo {title} {On the gravitational
  radiation from the collapse of neutron stars to rotating black holes},}\
  }\href {\doibase 10.1088/0264-9381/24/12/s13} {\bibfield  {journal} {\bibinfo
   {journal} {Classical and Quantum Gravity}\ }\textbf {\bibinfo {volume}
  {24}},\ \bibinfo {pages} {S187--S206} (\bibinfo {year} {2007})}\BibitemShut
  {NoStop}%
\bibitem [{\citenamefont {Damour}\ and\ \citenamefont
  {Vilenkin}(2000)}]{DV-2000}%
  \BibitemOpen
  \bibfield  {author} {\bibinfo {author} {\bibfnamefont {T.}~\bibnamefont
  {Damour}}\ and\ \bibinfo {author} {\bibfnamefont {A.}~\bibnamefont
  {Vilenkin}},\ }\bibfield  {title} {\enquote {\bibinfo {title} {Gravitational
  wave bursts from cosmic strings},}\ }\href {\doibase
  10.1103/PhysRevLett.85.3761} {\bibfield  {journal} {\bibinfo  {journal}
  {Phys. Rev. Lett.}\ }\textbf {\bibinfo {volume} {85}},\ \bibinfo {pages}
  {3761--3764} (\bibinfo {year} {2000})}\BibitemShut {NoStop}%
\bibitem [{\citenamefont {Damour}\ and\ \citenamefont
  {Vilenkin}(2001)}]{DV-2001}%
  \BibitemOpen
  \bibfield  {author} {\bibinfo {author} {\bibfnamefont {T.}~\bibnamefont
  {Damour}}\ and\ \bibinfo {author} {\bibfnamefont {A.}~\bibnamefont
  {Vilenkin}},\ }\bibfield  {title} {\enquote {\bibinfo {title} {Gravitational
  wave bursts from cusps and kinks on cosmic strings},}\ }\href {\doibase
  10.1103/PhysRevD.64.064008} {\bibfield  {journal} {\bibinfo  {journal} {Phys.
  Rev. D}\ }\textbf {\bibinfo {volume} {64}},\ \bibinfo {pages} {064008}
  (\bibinfo {year} {2001})}\BibitemShut {NoStop}%
\bibitem [{\citenamefont {Damour}\ and\ \citenamefont
  {Vilenkin}(2005)}]{PhysRevD.71.063510}%
  \BibitemOpen
  \bibfield  {author} {\bibinfo {author} {\bibfnamefont {T.}~\bibnamefont
  {Damour}}\ and\ \bibinfo {author} {\bibfnamefont {A.}~\bibnamefont
  {Vilenkin}},\ }\bibfield  {title} {\enquote {\bibinfo {title} {Gravitational
  radiation from cosmic (super)strings: Bursts, stochastic background, and
  observational windows},}\ }\href {\doibase 10.1103/PhysRevD.71.063510}
  {\bibfield  {journal} {\bibinfo  {journal} {Phys. Rev. D}\ }\textbf {\bibinfo
  {volume} {71}},\ \bibinfo {pages} {063510} (\bibinfo {year}
  {2005})}\BibitemShut {NoStop}%
\bibitem [{\citenamefont {Abbott}\ \emph
  {et~al.}(2017{\natexlab{b}})\citenamefont {Abbott} \emph
  {et~al.}}]{PhysRevD.96.022001}%
  \BibitemOpen
  \bibfield  {author} {\bibinfo {author} {\bibfnamefont {B.~P.}\ \bibnamefont
  {Abbott}} \emph {et~al.} (\bibinfo {collaboration} {LIGO Scientific
  Collaboration and Virgo Collaboration}),\ }\bibfield  {title} {\enquote
  {\bibinfo {title} {Search for intermediate mass black hole binaries in the
  first observing run of {A}dvanced {LIGO}},}\ }\href {\doibase
  10.1103/PhysRevD.96.022001} {\bibfield  {journal} {\bibinfo  {journal} {Phys.
  Rev. D}\ }\textbf {\bibinfo {volume} {96}},\ \bibinfo {pages} {022001}
  (\bibinfo {year} {2017}{\natexlab{b}})}\BibitemShut {NoStop}%
\bibitem [{\citenamefont {Abbott}\ \emph
  {et~al.}(2009{\natexlab{a}})\citenamefont {Abbott} \emph
  {et~al.}}]{Abbott:2009rr}%
  \BibitemOpen
  \bibfield  {author} {\bibinfo {author} {\bibfnamefont {B.~P.}\ \bibnamefont
  {Abbott}} \emph {et~al.} (\bibinfo {collaboration} {LIGO Scientific
  Collaboration}),\ }\bibfield  {title} {\enquote {\bibinfo {title} {{First
  LIGO search for gravitational wave bursts from cosmic (super)strings}},}\
  }\href {\doibase 10.1103/PhysRevD.80.062002} {\bibfield  {journal} {\bibinfo
  {journal} {Phys.Rev.}\ }\textbf {\bibinfo {volume} {D80}},\ \bibinfo {pages}
  {062002} (\bibinfo {year} {2009}{\natexlab{a}})},\ \Eprint
  {http://arxiv.org/abs/0904.4718} {arXiv:0904.4718 [astro-ph.CO]} \BibitemShut
  {NoStop}%
%%CITATION = ARXIV:0904.4718;%%
\bibitem [{\citenamefont {Aasi}\ \emph {et~al.}(2014)\citenamefont {Aasi} \emph
  {et~al.}}]{Aasi:2013vna}%
  \BibitemOpen
  \bibfield  {author} {\bibinfo {author} {\bibfnamefont {J.}~\bibnamefont
  {Aasi}} \emph {et~al.} (\bibinfo {collaboration} {LIGO Scientific
  Collaboration and Virgo Collaboration}),\ }\bibfield  {title} {\enquote
  {\bibinfo {title} {{Constraints on cosmic strings from the LIGO-Virgo
  gravitational-wave detectors}},}\ }\href {\doibase
  10.1103/PhysRevLett.112.131101} {\bibfield  {journal} {\bibinfo  {journal}
  {Phys.Rev.Lett.}\ }\textbf {\bibinfo {volume} {112}},\ \bibinfo {pages}
  {131101} (\bibinfo {year} {2014})},\ \Eprint {http://arxiv.org/abs/1310.2384}
  {arXiv:1310.2384 [gr-qc]} \BibitemShut {NoStop}%
%%CITATION = ARXIV:1310.2384;%%
\bibitem [{\citenamefont {Abbott}\ \emph
  {et~al.}(2018{\natexlab{b}})\citenamefont {Abbott} \emph
  {et~al.}}]{Abbott:2017mem}%
  \BibitemOpen
  \bibfield  {author} {\bibinfo {author} {\bibfnamefont {B.~P.}\ \bibnamefont
  {Abbott}} \emph {et~al.} (\bibinfo {collaboration} {LIGO Scientific
  Collaboration and Virgo Collaboration}),\ }\bibfield  {title} {\enquote
  {\bibinfo {title} {{Constraints on cosmic strings using data from the first
  Advanced LIGO observing run}},}\ }\href {\doibase 10.1103/PhysRevD.97.102002}
  {\bibfield  {journal} {\bibinfo  {journal} {Phys. Rev.}\ }\textbf {\bibinfo
  {volume} {D97}},\ \bibinfo {pages} {102002} (\bibinfo {year}
  {2018}{\natexlab{b}})},\ \Eprint {http://arxiv.org/abs/1712.01168}
  {arXiv:1712.01168 [gr-qc]} \BibitemShut {NoStop}%
%%CITATION = ARXIV:1712.01168;%%
\bibitem [{\citenamefont {Abbott}\ \emph {et~al.}(2015)\citenamefont {Abbott}
  \emph {et~al.}}]{Abbott:2015vir}%
  \BibitemOpen
  \bibfield  {author} {\bibinfo {author} {\bibfnamefont {B.~P.}\ \bibnamefont
  {Abbott}} \emph {et~al.} (\bibinfo {collaboration} {LIGO Scientific
  Collaboration and Virgo Collaboration}),\ }\bibfield  {title} {\enquote
  {\bibinfo {title} {{An all-sky search for long-duration gravitational wave
  transients with LIGO}},}\ }\href {\doibase 0.1088/1361-6382/aaab76}
  {\bibfield  {journal} {\bibinfo  {journal} {Phys.Rev. D}\ }\textbf {\bibinfo
  {volume} {93}},\ \bibinfo {pages} {042005} (\bibinfo {year} {2015})},\
  \Eprint {http://arxiv.org/abs/1511.04398} {arXiv:1511.04398 [gr-qc]}
  \BibitemShut {NoStop}%
%%CITATION = ARXIV:1511.04398;%%
\bibitem [{\citenamefont {Abbott}\ \emph
  {et~al.}(2018{\natexlab{c}})\citenamefont {Abbott} \emph
  {et~al.}}]{Abbott:2017muc}%
  \BibitemOpen
  \bibfield  {author} {\bibinfo {author} {\bibfnamefont {B.~P.}\ \bibnamefont
  {Abbott}} \emph {et~al.} (\bibinfo {collaboration} {LIGO Scientific
  Collaboration and Virgo Collaboration}),\ }\bibfield  {title} {\enquote
  {\bibinfo {title} {{All-sky search for long-duration gravitational wave
  transients in the first Advanced LIGO observing run}},}\ }\href {\doibase
  10.1088/1361-6382/aaab76} {\bibfield  {journal} {\bibinfo  {journal} {Class.
  Quant. Grav.}\ }\textbf {\bibinfo {volume} {35}},\ \bibinfo {pages} {065009}
  (\bibinfo {year} {2018}{\natexlab{c}})},\ \Eprint
  {http://arxiv.org/abs/1711.06843} {arXiv:1711.06843 [gr-qc]} \BibitemShut
  {NoStop}%
%%CITATION = ARXIV:1711.06843;%%
\bibitem [{\citenamefont {Abbott}\ \emph
  {et~al.}(2019{\natexlab{a}})\citenamefont {Abbott} \emph
  {et~al.}}]{Abbott:2019heg}%
  \BibitemOpen
  \bibfield  {author} {\bibinfo {author} {\bibfnamefont {B.~P.}\ \bibnamefont
  {Abbott}} \emph {et~al.} (\bibinfo {collaboration} {LIGO Scientific
  Collaboration and Virgo Collaboration}),\ }\bibfield  {title} {\enquote
  {\bibinfo {title} {{All-sky search for long-duration gravitational-wave
  transients in the second Advanced LIGO observing run}},}\ }\href@noop {} {\
  (\bibinfo {year} {2019}{\natexlab{a}})},\ \Eprint
  {http://arxiv.org/abs/1903.12015} {arXiv:1903.12015 [gr-qc]} \BibitemShut
  {NoStop}%
%%CITATION = ARXIV:1903.12015;%%
\bibitem [{\citenamefont {Abadie}\ \emph {et~al.}(2010)\citenamefont {Abadie}
  \emph {et~al.}}]{Abadie:2010mt}%
  \BibitemOpen
  \bibfield  {author} {\bibinfo {author} {\bibfnamefont {J.}~\bibnamefont
  {Abadie}} \emph {et~al.} (\bibinfo {collaboration} {LIGO Scientific
  Collaboration and Virgo Collaboration}),\ }\bibfield  {title} {\enquote
  {\bibinfo {title} {{All-sky search for gravitational-wave bursts in the first
  joint LIGO-GEO-Virgo run}},}\ }\href {\doibase 10.1103/PhysRevD.85.089905,
  10.1103/PhysRevD.81.102001} {\bibfield  {journal} {\bibinfo  {journal} {Phys.
  Rev.}\ }\textbf {\bibinfo {volume} {D81}},\ \bibinfo {pages} {102001}
  (\bibinfo {year} {2010})},\ \Eprint {http://arxiv.org/abs/1002.1036}
  {arXiv:1002.1036 [gr-qc]} \BibitemShut {NoStop}%
%%CITATION = ARXIV:1002.1036;%%
\bibitem [{\citenamefont {Abadie}\ \emph {et~al.}(2012)\citenamefont {Abadie}
  \emph {et~al.}}]{s6Burst}%
  \BibitemOpen
  \bibfield  {author} {\bibinfo {author} {\bibfnamefont {J.}~\bibnamefont
  {Abadie}} \emph {et~al.} (\bibinfo {collaboration} {LIGO Scientific
  Collaboration and Virgo Collaboration}),\ }\bibfield  {title} {\enquote
  {\bibinfo {title} {{All-sky search for gravitational-wave bursts in the
  second joint LIGO-Virgo run}},}\ }\href {\doibase 10.1103/PhysRevD.85.122007}
  {\bibfield  {journal} {\bibinfo  {journal} {\prd}\ }\textbf {\bibinfo
  {volume} {85}},\ \bibinfo {eid} {122007} (\bibinfo {year} {2012})},\ \Eprint
  {http://arxiv.org/abs/1202.2788} {arXiv:1202.2788 [gr-qc]} \BibitemShut
  {NoStop}%
\bibitem [{\citenamefont {Abbott}\ \emph
  {et~al.}(2017{\natexlab{c}})\citenamefont {Abbott} \emph
  {et~al.}}]{O1AllSky}%
  \BibitemOpen
  \bibfield  {author} {\bibinfo {author} {\bibfnamefont {B.~P.}\ \bibnamefont
  {Abbott}} \emph {et~al.} (\bibinfo {collaboration} {LIGO Scientific
  Collaboration and Virgo Collaboration}),\ }\bibfield  {title} {\enquote
  {\bibinfo {title} {All-sky search for short gravitational-wave bursts in the
  first advanced ligo run},}\ }\href {\doibase 10.1103/PhysRevD.95.042003}
  {\bibfield  {journal} {\bibinfo  {journal} {Phys. Rev. D}\ }\textbf {\bibinfo
  {volume} {95}},\ \bibinfo {pages} {042003} (\bibinfo {year}
  {2017}{\natexlab{c}})}\BibitemShut {NoStop}%
\bibitem [{\citenamefont {Abbott}\ \emph
  {et~al.}(2018{\natexlab{d}})\citenamefont {Abbott} \emph
  {et~al.}}]{ObservingScenario}%
  \BibitemOpen
  \bibfield  {author} {\bibinfo {author} {\bibfnamefont {B.~P.}\ \bibnamefont
  {Abbott}} \emph {et~al.} (\bibinfo {collaboration} {LIGO Scientific
  Collaboration and Virgo Collaboration}),\ }\bibfield  {title} {\enquote
  {\bibinfo {title} {Prospects for observing and localizing gravitational-wave
  transients with {A}dvanced {LIGO}, {A}dvanced {Virgo} and {KAGRA}s},}\ }\href
  {\doibase 10.1007/s41114-018-0012-9} {\bibfield  {journal} {\bibinfo
  {journal} {Living Reviews in Relativity}\ }\textbf {\bibinfo {volume} {21}},\
  \bibinfo {pages} {3} (\bibinfo {year} {2018}{\natexlab{d}})}\BibitemShut
  {NoStop}%
\bibitem [{\citenamefont {Abbott}\ \emph
  {et~al.}(2016{\natexlab{a}})\citenamefont {Abbott} \emph
  {et~al.}}]{DetcharCompanion}%
  \BibitemOpen
  \bibfield  {author} {\bibinfo {author} {\bibfnamefont {B.~P.}\ \bibnamefont
  {Abbott}} \emph {et~al.} (\bibinfo {collaboration} {LIGO Scientific
  Collaboration and Virgo Collaboration}),\ }\bibfield  {title} {\enquote
  {\bibinfo {title} {{Characterization of transient noise in Advanced LIGO
  relevant to gravitational wave signal GW150914}},}\ }\href {\doibase
  10.1088/0264-9381/33/13/134001} {\bibfield  {journal} {\bibinfo  {journal}
  {Class. Quant. Grav.}\ }\textbf {\bibinfo {volume} {33}},\ \bibinfo {pages}
  {134001} (\bibinfo {year} {2016}{\natexlab{a}})},\ \Eprint
  {http://arxiv.org/abs/1602.03844} {arXiv:1602.03844 [gr-qc]} \BibitemShut
  {NoStop}%
%%CITATION = ARXIV:1602.03844;%%
\bibitem [{\citenamefont {Abbott}\ \emph
  {et~al.}(2016{\natexlab{b}})\citenamefont {Abbott} \emph
  {et~al.}}]{BurstCompanion}%
  \BibitemOpen
  \bibfield  {author} {\bibinfo {author} {\bibfnamefont {B.~P.}\ \bibnamefont
  {Abbott}} \emph {et~al.} (\bibinfo {collaboration} {LIGO Scientific
  Collaboration and Virgo Collaboration}),\ }\bibfield  {title} {\enquote
  {\bibinfo {title} {{Observing gravitational-wave transient GW150914 with
  minimal assumptions}},}\ }\href {\doibase 10.1103/PhysRevD.94.069903,
  10.1103/PhysRevD.93.122004} {\bibfield  {journal} {\bibinfo  {journal} {Phys.
  Rev.}\ }\textbf {\bibinfo {volume} {D93}},\ \bibinfo {pages} {122004}
  (\bibinfo {year} {2016}{\natexlab{b}})},\ \bibinfo {note} {[Addendum: Phys.
  Rev.D94,no.6,069903(2016)]},\ \Eprint {http://arxiv.org/abs/1602.03843}
  {arXiv:1602.03843 [gr-qc]} \BibitemShut {NoStop}%
%%CITATION = ARXIV:1602.03843;%%
\bibitem [{\citenamefont {Abbott}\ \emph
  {et~al.}(2018{\natexlab{e}})\citenamefont {Abbott} \emph
  {et~al.}}]{O1-Detchar}%
  \BibitemOpen
  \bibfield  {author} {\bibinfo {author} {\bibfnamefont {B.~P.}\ \bibnamefont
  {Abbott}} \emph {et~al.} (\bibinfo {collaboration} {LIGO Scientific
  Collaboration and Virgo Collaboration}),\ }\bibfield  {title} {\enquote
  {\bibinfo {title} {{Effects of data quality vetoes on a search for compact
  binary coalescences in Advanced LIGO first observing run}},}\ }\href
  {\doibase 10.1088/1361-6382/aaaafa} {\bibfield  {journal} {\bibinfo
  {journal} {Class. Quant. Grav.}\ }\textbf {\bibinfo {volume} {35}},\ \bibinfo
  {pages} {065010} (\bibinfo {year} {2018}{\natexlab{e}})},\ \Eprint
  {http://arxiv.org/abs/1710.02185} {arXiv:1710.02185 [gr-qc]} \BibitemShut
  {NoStop}%
%%CITATION = ARXIV:1710.02185;%%
\bibitem [{\citenamefont {Nuttall}(2018)}]{Nuttall:2018xhi}%
  \BibitemOpen
  \bibfield  {author} {\bibinfo {author} {\bibfnamefont {L.~K.}\ \bibnamefont
  {Nuttall}},\ }\bibfield  {title} {\enquote {\bibinfo {title} {{Characterizing
  transient noise in the LIGO detectors}},}\ }\href {\doibase
  10.1098/rsta.2017.0286} {\bibfield  {journal} {\bibinfo  {journal} {Phil.
  Trans. Roy. Soc. Lond.}\ }\textbf {\bibinfo {volume} {A376}},\ \bibinfo
  {pages} {20170286} (\bibinfo {year} {2018})},\ \Eprint
  {http://arxiv.org/abs/1804.07592} {arXiv:1804.07592 [astro-ph.IM]}
  \BibitemShut {NoStop}%
%%CITATION = ARXIV:1804.07592;%%
\bibitem [{\citenamefont {Viets}\ \emph {et~al.}(2018)\citenamefont {Viets},
  \citenamefont {Wade}, \citenamefont {Urban}, \citenamefont {Kandhasamy},
  \citenamefont {Betzwieser}, \citenamefont {Brown}, \citenamefont
  {Burguet-Castell}, \citenamefont {Cahillane}, \citenamefont {Goetz},
  \citenamefont {Izumi}, \citenamefont {Karki}, \citenamefont {Kissel},
  \citenamefont {Mendell}, \citenamefont {Savage}, \citenamefont {Siemens},
  \citenamefont {Tuyenbayev},\ and\ \citenamefont {Weinstein}}]{Viets_2018}%
  \BibitemOpen
  \bibfield  {author} {\bibinfo {author} {\bibfnamefont {A.~D.}\ \bibnamefont
  {Viets}}, \bibinfo {author} {\bibfnamefont {M.}~\bibnamefont {Wade}},
  \bibinfo {author} {\bibfnamefont {A.~L.}\ \bibnamefont {Urban}}, \bibinfo
  {author} {\bibfnamefont {S.}~\bibnamefont {Kandhasamy}}, \bibinfo {author}
  {\bibfnamefont {J.}~\bibnamefont {Betzwieser}}, \bibinfo {author}
  {\bibfnamefont {D.~A.}\ \bibnamefont {Brown}}, \bibinfo {author}
  {\bibfnamefont {J.}~\bibnamefont {Burguet-Castell}}, \bibinfo {author}
  {\bibfnamefont {C.}~\bibnamefont {Cahillane}}, \bibinfo {author}
  {\bibfnamefont {E.}~\bibnamefont {Goetz}}, \bibinfo {author} {\bibfnamefont
  {K.}~\bibnamefont {Izumi}}, \bibinfo {author} {\bibfnamefont
  {S.}~\bibnamefont {Karki}}, \bibinfo {author} {\bibfnamefont {J.~S.}\
  \bibnamefont {Kissel}}, \bibinfo {author} {\bibfnamefont {G.}~\bibnamefont
  {Mendell}}, \bibinfo {author} {\bibfnamefont {R.~L.}\ \bibnamefont {Savage}},
  \bibinfo {author} {\bibfnamefont {X.}~\bibnamefont {Siemens}}, \bibinfo
  {author} {\bibfnamefont {D.}~\bibnamefont {Tuyenbayev}}, \ and\ \bibinfo
  {author} {\bibfnamefont {A.~J.}\ \bibnamefont {Weinstein}},\ }\bibfield
  {title} {\enquote {\bibinfo {title} {Reconstructing the calibrated strain
  signal in the advanced {LIGO} detectors},}\ }\href {\doibase
  10.1088/1361-6382/aab658} {\bibfield  {journal} {\bibinfo  {journal}
  {Classical and Quantum Gravity}\ }\textbf {\bibinfo {volume} {35}},\ \bibinfo
  {pages} {095015} (\bibinfo {year} {2018})}\BibitemShut {NoStop}%
\bibitem [{\citenamefont {Cahillane}\ \emph {et~al.}(2017)\citenamefont
  {Cahillane}, \citenamefont {Betzwieser}, \citenamefont {Brown}, \citenamefont
  {Goetz}, \citenamefont {Hall}, \citenamefont {Izumi}, \citenamefont
  {Kandhasamy}, \citenamefont {Karki}, \citenamefont {Kissel}, \citenamefont
  {Mendell}, \citenamefont {Savage}, \citenamefont {Tuyenbayev}, \citenamefont
  {Urban}, \citenamefont {Viets}, \citenamefont {Wade},\ and\ \citenamefont
  {Weinstein}}]{PhysRevD.96.102001}%
  \BibitemOpen
  \bibfield  {author} {\bibinfo {author} {\bibfnamefont {C.}~\bibnamefont
  {Cahillane}}, \bibinfo {author} {\bibfnamefont {J.}~\bibnamefont
  {Betzwieser}}, \bibinfo {author} {\bibfnamefont {D.~A.}\ \bibnamefont
  {Brown}}, \bibinfo {author} {\bibfnamefont {E.}~\bibnamefont {Goetz}},
  \bibinfo {author} {\bibfnamefont {E.~D.}\ \bibnamefont {Hall}}, \bibinfo
  {author} {\bibfnamefont {K.}~\bibnamefont {Izumi}}, \bibinfo {author}
  {\bibfnamefont {S.}~\bibnamefont {Kandhasamy}}, \bibinfo {author}
  {\bibfnamefont {S.}~\bibnamefont {Karki}}, \bibinfo {author} {\bibfnamefont
  {J.~S.}\ \bibnamefont {Kissel}}, \bibinfo {author} {\bibfnamefont
  {G.}~\bibnamefont {Mendell}}, \bibinfo {author} {\bibfnamefont {R.~L.}\
  \bibnamefont {Savage}}, \bibinfo {author} {\bibfnamefont {D.}~\bibnamefont
  {Tuyenbayev}}, \bibinfo {author} {\bibfnamefont {A.}~\bibnamefont {Urban}},
  \bibinfo {author} {\bibfnamefont {A.}~\bibnamefont {Viets}}, \bibinfo
  {author} {\bibfnamefont {M.}~\bibnamefont {Wade}}, \ and\ \bibinfo {author}
  {\bibfnamefont {A.~J.}\ \bibnamefont {Weinstein}},\ }\bibfield  {title}
  {\enquote {\bibinfo {title} {Calibration uncertainty for advanced ligo's
  first and second observing runs},}\ }\href {\doibase
  10.1103/PhysRevD.96.102001} {\bibfield  {journal} {\bibinfo  {journal} {Phys.
  Rev. D}\ }\textbf {\bibinfo {volume} {96}},\ \bibinfo {pages} {102001}
  (\bibinfo {year} {2017})}\BibitemShut {NoStop}%
\bibitem [{\citenamefont {Driggers}\ \emph {et~al.}(2019)\citenamefont
  {Driggers} \emph {et~al.}}]{Driggers:2018gii}%
  \BibitemOpen
  \bibfield  {author} {\bibinfo {author} {\bibfnamefont {J.~C.}\ \bibnamefont
  {Driggers}} \emph {et~al.} (\bibinfo {collaboration} {LIGO Scientific
  Collaboration}),\ }\bibfield  {title} {\enquote {\bibinfo {title} {{Improving
  astrophysical parameter estimation via offline noise subtraction for Advanced
  LIGO}},}\ }\href {\doibase 10.1103/PhysRevD.99.042001} {\bibfield  {journal}
  {\bibinfo  {journal} {Phys. Rev.}\ }\textbf {\bibinfo {volume} {D99}},\
  \bibinfo {pages} {042001} (\bibinfo {year} {2019})},\ \Eprint
  {http://arxiv.org/abs/1806.00532} {arXiv:1806.00532 [astro-ph.IM]}
  \BibitemShut {NoStop}%
\bibitem [{\citenamefont {Vallisneri}\ \emph {et~al.}(2015)\citenamefont
  {Vallisneri}, \citenamefont {Kanner}, \citenamefont {Williams}, \citenamefont
  {Weinstein},\ and\ \citenamefont {Stephens}}]{Vallisneri_2015}%
  \BibitemOpen
  \bibfield  {author} {\bibinfo {author} {\bibfnamefont {M.}~\bibnamefont
  {Vallisneri}}, \bibinfo {author} {\bibfnamefont {J.}~\bibnamefont {Kanner}},
  \bibinfo {author} {\bibfnamefont {R.}~\bibnamefont {Williams}}, \bibinfo
  {author} {\bibfnamefont {A.}~\bibnamefont {Weinstein}}, \ and\ \bibinfo
  {author} {\bibfnamefont {B.}~\bibnamefont {Stephens}},\ }\bibfield  {title}
  {\enquote {\bibinfo {title} {The {LIGO} open science center},}\ }\href
  {\doibase 10.1088/1742-6596/610/1/012021} {\bibfield  {journal} {\bibinfo
  {journal} {Journal of Physics: Conference Series}\ }\textbf {\bibinfo
  {volume} {610}},\ \bibinfo {pages} {012021} (\bibinfo {year}
  {2015})}\BibitemShut {NoStop}%
\bibitem [{\citenamefont {{LIGO Scientific Collaboration And Virgo
  Collaboration}}(2019)}]{GWOSC}%
  \BibitemOpen
  \bibfield  {author} {\bibinfo {author} {\bibnamefont {{LIGO Scientific
  Collaboration And Virgo Collaboration}}},\ }\href {\doibase
  10.7935/ca75-fm95} {\enquote {\bibinfo {title} {{LIGO} {V}irgo strain data
  from observing run {O}2 ({N}ov 2016 - {A}ug 2017)},}\ }\bibinfo
  {howpublished} {\url{https://www.gw-openscience.org/O2/}} (\bibinfo {year}
  {2019})\BibitemShut {NoStop}%
\bibitem [{\citenamefont {Usman}\ \emph {et~al.}(2016)\citenamefont {Usman},
  \citenamefont {.Nitz}, \citenamefont {Harry}, \citenamefont {Biwer},
  \citenamefont {Brown}, \citenamefont {Cabero}, \citenamefont {Capano},
  \citenamefont {Canton}, \citenamefont {Dent}, \citenamefont {Fairhurst},
  \citenamefont {Kehl}, \citenamefont {Keppel}, \citenamefont {Krishnan},
  \citenamefont {Lenon}, \citenamefont {Lundgren}, \citenamefont {Nielsen},
  \citenamefont {L.~P}, \citenamefont {Pfeiffer}, \citenamefont {Saulson},
  \citenamefont {West},\ and\ \citenamefont {Willis}}]{Usman_2016}%
  \BibitemOpen
  \bibfield  {author} {\bibinfo {author} {\bibfnamefont {S.~A.}\ \bibnamefont
  {Usman}}, \bibinfo {author} {\bibfnamefont {A.~H}\ \bibnamefont {.Nitz}},
  \bibinfo {author} {\bibfnamefont {I.~W.}\ \bibnamefont {Harry}}, \bibinfo
  {author} {\bibfnamefont {C.~M.}\ \bibnamefont {Biwer}}, \bibinfo {author}
  {\bibfnamefont {D.~A.}\ \bibnamefont {Brown}}, \bibinfo {author}
  {\bibfnamefont {M.}~\bibnamefont {Cabero}}, \bibinfo {author} {\bibfnamefont
  {C.~D.}\ \bibnamefont {Capano}}, \bibinfo {author} {\bibfnamefont {T.~Dal}\
  \bibnamefont {Canton}}, \bibinfo {author} {\bibfnamefont {T.}~\bibnamefont
  {Dent}}, \bibinfo {author} {\bibfnamefont {S.}~\bibnamefont {Fairhurst}},
  \bibinfo {author} {\bibfnamefont {M.~S.}\ \bibnamefont {Kehl}}, \bibinfo
  {author} {\bibfnamefont {D.}~\bibnamefont {Keppel}}, \bibinfo {author}
  {\bibfnamefont {B.}~\bibnamefont {Krishnan}}, \bibinfo {author}
  {\bibfnamefont {A.}~\bibnamefont {Lenon}}, \bibinfo {author} {\bibfnamefont
  {A.}~\bibnamefont {Lundgren}}, \bibinfo {author} {\bibfnamefont {A.~B.}\
  \bibnamefont {Nielsen}}, \bibinfo {author} {\bibfnamefont {Pekowsky}\
  \bibnamefont {L.~P}}, \bibinfo {author} {\bibfnamefont {H.~P.}\ \bibnamefont
  {Pfeiffer}}, \bibinfo {author} {\bibfnamefont {P.~R.}\ \bibnamefont
  {Saulson}}, \bibinfo {author} {\bibfnamefont {M.}~\bibnamefont {West}}, \
  and\ \bibinfo {author} {\bibfnamefont {J.~L}\ \bibnamefont {Willis}},\
  }\bibfield  {title} {\enquote {\bibinfo {title} {The {PyCBC} search for
  gravitational waves from compact binary coalescence},}\ }\href {\doibase
  10.1088/0264-9381/33/21/215004} {\bibfield  {journal} {\bibinfo  {journal}
  {Classical and Quantum Gravity}\ }\textbf {\bibinfo {volume} {33}},\ \bibinfo
  {pages} {215004} (\bibinfo {year} {2016})}\BibitemShut {NoStop}%
\bibitem [{\citenamefont {Nitz}\ \emph {et~al.}(2017)\citenamefont {Nitz} \emph
  {et~al.}}]{alex_nitz_2017_344823}%
  \BibitemOpen
  \bibfield  {author} {\bibinfo {author} {\bibfnamefont {A.}~\bibnamefont
  {Nitz}} \emph {et~al.},\ }\href {\doibase 10.5281/zenodo.344823} {\enquote
  {\bibinfo {title} {Pycbc: O2 production release 9},}\ }\bibinfo
  {howpublished} {free software,
  \url{https://github.com/gwastro/pycbc/tree/v1.6.8}} (\bibinfo {year}
  {2017})\BibitemShut {NoStop}%
\bibitem [{\citenamefont {Messick}(2017)}]{PhysRevD.95.042001}%
  \BibitemOpen
  \bibfield  {author} {\bibinfo {author} {\bibfnamefont {C.}~\bibnamefont
  {Messick}},\ }\bibfield  {title} {\enquote {\bibinfo {title} {Analysis
  framework for the prompt discovery of compact binary mergers in
  gravitational-wave data},}\ }\href {\doibase 10.1103/PhysRevD.95.042001}
  {\bibfield  {journal} {\bibinfo  {journal} {Phys. Rev. D}\ }\textbf {\bibinfo
  {volume} {95}},\ \bibinfo {pages} {042001} (\bibinfo {year}
  {2017})}\BibitemShut {NoStop}%
\bibitem [{\citenamefont {Klimenko}\ \emph {et~al.}(2015)\citenamefont
  {Klimenko}, \citenamefont {Vedovato}, \citenamefont {Drago}, \citenamefont
  {Salemi}, \citenamefont {Tiwari}, \citenamefont {Prodi}, \citenamefont
  {Lazzaro}, \citenamefont {Tiwari}, \citenamefont {Da~Silva},\ and\
  \citenamefont {Mitselmakher}}]{cwb2g}%
  \BibitemOpen
  \bibfield  {author} {\bibinfo {author} {\bibfnamefont {S.}~\bibnamefont
  {Klimenko}}, \bibinfo {author} {\bibfnamefont {G.}~\bibnamefont {Vedovato}},
  \bibinfo {author} {\bibfnamefont {M.}~\bibnamefont {Drago}}, \bibinfo
  {author} {\bibfnamefont {F.}~\bibnamefont {Salemi}}, \bibinfo {author}
  {\bibfnamefont {V.}~\bibnamefont {Tiwari}}, \bibinfo {author} {\bibfnamefont
  {G.~A.}\ \bibnamefont {Prodi}}, \bibinfo {author} {\bibfnamefont
  {C.}~\bibnamefont {Lazzaro}}, \bibinfo {author} {\bibfnamefont
  {S.}~\bibnamefont {Tiwari}}, \bibinfo {author} {\bibfnamefont
  {F.}~\bibnamefont {Da~Silva}}, \ and\ \bibinfo {author} {\bibfnamefont
  {G.}~\bibnamefont {Mitselmakher}},\ }\bibfield  {title} {\enquote {\bibinfo
  {title} {{Method for detection and reconstruction of gravitational wave
  transients with networks of advanced detectors}},}\ }\href {\doibase
  10.1103/PhysRevD.93.042004} {\bibfield  {journal} {\bibinfo  {journal} {Phys.
  Rev. D 93, 042004}\ } (\bibinfo {year} {2015}),\
  10.1103/PhysRevD.93.042004},\ \Eprint {http://arxiv.org/abs/1511.05999}
  {arXiv:1511.05999 [gr-qc]} \BibitemShut {NoStop}%
%%CITATION = ARXIV:1511.05999;%%
\bibitem [{gli()}]{glitch_alog}%
  \BibitemOpen
  \href@noop {} {}\bibinfo {howpublished}
  {\url{https://alog.ligo-wa.caltech.edu/aLOG/index.php?callRep=33104}}\BibitemShut
  {NoStop}%
\bibitem [{\citenamefont {Abbott}\ \emph
  {et~al.}(2009{\natexlab{b}})\citenamefont {Abbott} \emph {et~al.}}]{S5burst}%
  \BibitemOpen
  \bibfield  {author} {\bibinfo {author} {\bibfnamefont {B.~P.}\ \bibnamefont
  {Abbott}} \emph {et~al.} (\bibinfo {collaboration} {LIGO Scientific
  Collaboration}),\ }\bibfield  {title} {\enquote {\bibinfo {title} {{Search
  for gravitational-wave bursts in the first year of the fifth LIGO science
  run}},}\ }\href {\doibase 10.1103/PhysRevD.80.102001} {\bibfield  {journal}
  {\bibinfo  {journal} {Phys. Rev.}\ }\textbf {\bibinfo {volume} {D80}},\
  \bibinfo {pages} {102001} (\bibinfo {year} {2009}{\natexlab{b}})},\ \Eprint
  {http://arxiv.org/abs/0905.0020} {arXiv:0905.0020 [gr-qc]} \BibitemShut
  {NoStop}%
%%CITATION = ARXIV:0905.0020;%%
\bibitem [{\citenamefont {Robinet}(2015)}]{OmicronRobinet}%
  \BibitemOpen
  \bibfield  {author} {\bibinfo {author} {\bibfnamefont {F.}~\bibnamefont
  {Robinet}},\ }\href@noop {} {\enquote {\bibinfo {title} {{Omicron: an
  algorithm to detect and characterize transient noise in gravitational-wave
  detectors}},}\ }\bibinfo {howpublished}
  {{\url{https://tds.ego-gw.it/ql/?c=10651}}} (\bibinfo {year}
  {2015})\BibitemShut {NoStop}%
\bibitem [{\citenamefont {Lynch}\ \emph {et~al.}(2017)\citenamefont {Lynch},
  \citenamefont {Vitale}, \citenamefont {Essick}, \citenamefont
  {Katsavounidis},\ and\ \citenamefont {Robinet}}]{oLIBprep}%
  \BibitemOpen
  \bibfield  {author} {\bibinfo {author} {\bibfnamefont {R.}~\bibnamefont
  {Lynch}}, \bibinfo {author} {\bibfnamefont {S.}~\bibnamefont {Vitale}},
  \bibinfo {author} {\bibfnamefont {R.}~\bibnamefont {Essick}}, \bibinfo
  {author} {\bibfnamefont {E.}~\bibnamefont {Katsavounidis}}, \ and\ \bibinfo
  {author} {\bibfnamefont {F.}~\bibnamefont {Robinet}},\ }\bibfield  {title}
  {\enquote {\bibinfo {title} {Information-theoretic approach to the
  gravitational-wave burst detection problem},}\ }\href {\doibase
  10.1103/PhysRevD.95.104046} {\bibfield  {journal} {\bibinfo  {journal} {Phys.
  Rev. D}\ }\textbf {\bibinfo {volume} {95}},\ \bibinfo {pages} {104046}
  (\bibinfo {year} {2017})}\BibitemShut {NoStop}%
\bibitem [{\citenamefont {Lynch}\ \emph {et~al.}(2018)\citenamefont {Lynch},
  \citenamefont {Vitale},\ and\ \citenamefont {E.}}]{EventStacking}%
  \BibitemOpen
  \bibfield  {author} {\bibinfo {author} {\bibfnamefont {R.}~\bibnamefont
  {Lynch}}, \bibinfo {author} {\bibfnamefont {S.}~\bibnamefont {Vitale}}, \
  and\ \bibinfo {author} {\bibfnamefont {Katsavounidis}\ \bibnamefont {E.}},\
  }\bibfield  {title} {\enquote {\bibinfo {title} {{Beyond single-threshold
  searches: the Event Stacking Test}},}\ }\href@noop {} {\bibfield  {journal}
  {\bibinfo  {journal} {arXiv:1811.01297}\ } (\bibinfo {year} {2018})},\
  \Eprint {http://arxiv.org/abs/1811.01297} {arXiv:1811.01297
  [physics.data-an]} \BibitemShut {NoStop}%
%%CITATION = ARXIV:1511.05955;%%
\bibitem [{\citenamefont {Cornish}\ and\ \citenamefont
  {Littenberg}(2015)}]{Cornish:2014kda}%
  \BibitemOpen
  \bibfield  {author} {\bibinfo {author} {\bibfnamefont {N.J.}\ \bibnamefont
  {Cornish}}\ and\ \bibinfo {author} {\bibfnamefont {T.B.}\ \bibnamefont
  {Littenberg}},\ }\bibfield  {title} {\enquote {\bibinfo {title} {{BayesWave:
  Bayesian Inference for Gravitational Wave Bursts and Instrument Glitches}},}\
  }\href {\doibase 10.1088/0264-9381/32/13/135012} {\bibfield  {journal}
  {\bibinfo  {journal} {Class. Quant. Grav.}\ }\textbf {\bibinfo {volume}
  {32}},\ \bibinfo {pages} {135012} (\bibinfo {year} {2015})},\ \Eprint
  {http://arxiv.org/abs/1410.3835} {arXiv:1410.3835 [gr-qc]} \BibitemShut
  {NoStop}%
%%CITATION = ARXIV:1410.3835;%%
\bibitem [{bw_()}]{bw_repo}%
  \BibitemOpen
  \href@noop {} {}\bibinfo {howpublished}
  {\url{https://git.ligo.org/lscsoft/bayeswave}}\BibitemShut {NoStop}%
\bibitem [{\citenamefont {Kanner}\ \emph {et~al.}(2016)\citenamefont {Kanner},
  \citenamefont {Littenberg}, \citenamefont {Cornish}, \citenamefont
  {Millhouse}, \citenamefont {Xhakaj}, \citenamefont {Salemi}, \citenamefont
  {Drago}, \citenamefont {Vedovato},\ and\ \citenamefont
  {Klimenko}}]{Kanner:2015xua}%
  \BibitemOpen
  \bibfield  {author} {\bibinfo {author} {\bibfnamefont {J.~B.}\ \bibnamefont
  {Kanner}}, \bibinfo {author} {\bibfnamefont {T.~B.}\ \bibnamefont
  {Littenberg}}, \bibinfo {author} {\bibfnamefont {N.}~\bibnamefont {Cornish}},
  \bibinfo {author} {\bibfnamefont {M.}~\bibnamefont {Millhouse}}, \bibinfo
  {author} {\bibfnamefont {E.}~\bibnamefont {Xhakaj}}, \bibinfo {author}
  {\bibfnamefont {F.}~\bibnamefont {Salemi}}, \bibinfo {author} {\bibfnamefont
  {M.}~\bibnamefont {Drago}}, \bibinfo {author} {\bibfnamefont
  {G.}~\bibnamefont {Vedovato}}, \ and\ \bibinfo {author} {\bibfnamefont
  {S.}~\bibnamefont {Klimenko}},\ }\bibfield  {title} {\enquote {\bibinfo
  {title} {{Leveraging waveform complexity for confident detection of
  gravitational waves}},}\ }\href {\doibase 10.1103/PhysRevD.93.022002}
  {\bibfield  {journal} {\bibinfo  {journal} {Phys. Rev.}\ }\textbf {\bibinfo
  {volume} {D93}},\ \bibinfo {pages} {022002} (\bibinfo {year} {2016})},\
  \Eprint {http://arxiv.org/abs/1509.06423} {arXiv:1509.06423 [astro-ph.IM]}
  \BibitemShut {NoStop}%
%%CITATION = ARXIV:1509.06423;%%
\bibitem [{\citenamefont {Williams}(2018)}]{daniel_williams_2018_1699336}%
  \BibitemOpen
  \bibfield  {author} {\bibinfo {author} {\bibfnamefont {D.}~\bibnamefont
  {Williams}},\ }\href {\doibase 10.5281/zenodo.1699336} {\enquote {\bibinfo
  {title} {minke},}\ }\bibinfo {howpublished} {free software,
  \url{https://doi.org/10.5281/zenodo.1699336}} (\bibinfo {year}
  {2018})\BibitemShut {NoStop}%
\bibitem [{\citenamefont {Collaboration}(2018)}]{lalsuite}%
  \BibitemOpen
  \bibfield  {author} {\bibinfo {author} {\bibfnamefont {LIGO~Scientific}\
  \bibnamefont {Collaboration}},\ }\href {\doibase 10.7935/GT1W-FZ16} {\enquote
  {\bibinfo {title} {{LIGO} {A}lgorithm {L}ibrary - {LALS}uite},}\ }\bibinfo
  {howpublished} {free software (GPL),
  \url{https://git.ligo.org/lscsoft/lalsuite}} (\bibinfo {year}
  {2018})\BibitemShut {NoStop}%
\bibitem [{\citenamefont {Littenberg}\ \emph {et~al.}(2016)\citenamefont
  {Littenberg}, \citenamefont {Kanner}, \citenamefont {Cornish},\ and\
  \citenamefont {Millhouse}}]{Littenberg:2015kpb}%
  \BibitemOpen
  \bibfield  {author} {\bibinfo {author} {\bibfnamefont {T.B.}\ \bibnamefont
  {Littenberg}}, \bibinfo {author} {\bibfnamefont {J.B.}\ \bibnamefont
  {Kanner}}, \bibinfo {author} {\bibfnamefont {N.J.}\ \bibnamefont {Cornish}},
  \ and\ \bibinfo {author} {\bibfnamefont {M.}~\bibnamefont {Millhouse}},\
  }\bibfield  {title} {\enquote {\bibinfo {title} {{Enabling high confidence
  detections of gravitational-wave bursts}},}\ }\href {\doibase
  10.1103/PhysRevD.94.044050} {\bibfield  {journal} {\bibinfo  {journal} {Phys.
  Rev.}\ }\textbf {\bibinfo {volume} {D94}},\ \bibinfo {pages} {044050}
  (\bibinfo {year} {2016})},\ \Eprint {http://arxiv.org/abs/1511.08752}
  {arXiv:1511.08752} \BibitemShut {NoStop}%
\bibitem [{\citenamefont {{Vilenkin}}\ and\ \citenamefont
  {{Shellard}}(2000)}]{vilenkin-shellard}%
  \BibitemOpen
  \bibfield  {author} {\bibinfo {author} {\bibfnamefont {A.}~\bibnamefont
  {{Vilenkin}}}\ and\ \bibinfo {author} {\bibfnamefont {E.}~\bibnamefont
  {{Shellard}}},\ }\href@noop {} {\emph {\bibinfo {title} {{Cosmic strings and
  other Topological Defects}}}}\ (\bibinfo  {publisher} {Cambridge University
  Press},\ \bibinfo {year} {2000})\BibitemShut {NoStop}%
\bibitem [{\citenamefont {Goto}(1971)}]{Goto:1971ce}%
  \BibitemOpen
  \bibfield  {author} {\bibinfo {author} {\bibfnamefont {T.}~\bibnamefont
  {Goto}},\ }\bibfield  {title} {\enquote {\bibinfo {title} {{Relativistic
  quantum mechanics of one-dimensional mechanical continuum and subsidiary
  condition of dual resonance model}},}\ }\href {\doibase 10.1143/PTP.46.1560}
  {\bibfield  {journal} {\bibinfo  {journal} {Prog. Theor. Phys.}\ }\textbf
  {\bibinfo {volume} {46}},\ \bibinfo {pages} {1560--1569} (\bibinfo {year}
  {1971})}\BibitemShut {NoStop}%
%%CITATION = PTPKA,46,1560;%%
\bibitem [{\citenamefont {Cannon}(2008)}]{Cannon:2008zz}%
  \BibitemOpen
  \bibfield  {author} {\bibinfo {author} {\bibfnamefont {Kipp~C.}\ \bibnamefont
  {Cannon}},\ }\bibfield  {title} {\enquote {\bibinfo {title} {{A Bayesian
  coincidence test for noise rejection in a gravitational-wave burst
  search}},}\ }\href {\doibase 10.1088/0264-9381/25/10/105024} {\bibfield
  {journal} {\bibinfo  {journal} {Class.Quant.Grav.}\ }\textbf {\bibinfo
  {volume} {25}},\ \bibinfo {pages} {105024} (\bibinfo {year}
  {2008})}\BibitemShut {NoStop}%
\bibitem [{\citenamefont {Blanco-Pillado}\ \emph {et~al.}(2014)\citenamefont
  {Blanco-Pillado}, \citenamefont {Olum},\ and\ \citenamefont
  {Shlaer}}]{Blanco-Pillado:2013qja}%
  \BibitemOpen
  \bibfield  {author} {\bibinfo {author} {\bibfnamefont {J.~J.}\ \bibnamefont
  {Blanco-Pillado}}, \bibinfo {author} {\bibfnamefont {K.~D.}\ \bibnamefont
  {Olum}}, \ and\ \bibinfo {author} {\bibfnamefont {B.}~\bibnamefont
  {Shlaer}},\ }\bibfield  {title} {\enquote {\bibinfo {title} {{The number of
  cosmic string loops}},}\ }\href {\doibase 10.1103/PhysRevD.89.023512}
  {\bibfield  {journal} {\bibinfo  {journal} {Phys.Rev.}\ }\textbf {\bibinfo
  {volume} {D89}},\ \bibinfo {pages} {023512} (\bibinfo {year} {2014})},\
  \Eprint {http://arxiv.org/abs/1309.6637} {arXiv:1309.6637 [astro-ph.CO]}
  \BibitemShut {NoStop}%
%%CITATION = ARXIV:1309.6637;%%
\bibitem [{\citenamefont {Ringeval}\ \emph {et~al.}(2007)\citenamefont
  {Ringeval}, \citenamefont {Sakellariadou},\ and\ \citenamefont
  {Bouchet}}]{Ringeval:2005kr}%
  \BibitemOpen
  \bibfield  {author} {\bibinfo {author} {\bibfnamefont {C.}~\bibnamefont
  {Ringeval}}, \bibinfo {author} {\bibfnamefont {M.}~\bibnamefont
  {Sakellariadou}}, \ and\ \bibinfo {author} {\bibfnamefont {F.}~\bibnamefont
  {Bouchet}},\ }\bibfield  {title} {\enquote {\bibinfo {title} {{Cosmological
  evolution of cosmic string loops}},}\ }\href {\doibase
  10.1088/1475-7516/2007/02/023} {\bibfield  {journal} {\bibinfo  {journal}
  {JCAP}\ }\textbf {\bibinfo {volume} {0702}},\ \bibinfo {pages} {023}
  (\bibinfo {year} {2007})},\ \Eprint {http://arxiv.org/abs/astro-ph/0511646}
  {arXiv:astro-ph/0511646 [astro-ph]} \BibitemShut {NoStop}%
%%CITATION = ASTRO-PH/0511646;%%
\bibitem [{\citenamefont {Abbott}\ \emph
  {et~al.}(2019{\natexlab{b}})\citenamefont {Abbott} \emph
  {et~al.}}]{O2stochastic}%
  \BibitemOpen
  \bibfield  {author} {\bibinfo {author} {\bibfnamefont {B.~P.}\ \bibnamefont
  {Abbott}} \emph {et~al.} (\bibinfo {collaboration} {LIGO Scientific
  Collaboration and Virgo Collaboration}),\ }\bibfield  {title} {\enquote
  {\bibinfo {title} {{A search for the isotropic stochastic background using
  data from Advanced LIGO's second observing run}},}\ }\href@noop {} {\
  (\bibinfo {year} {2019}{\natexlab{b}})},\ \Eprint
  {http://arxiv.org/abs/1903.02886} {arXiv:1903.02886 [gr-qc]} \BibitemShut
  {NoStop}%
\end{thebibliography}
